\documentclass[useAMS,usenatbib]{mn2e}

\usepackage{amsmath,array,amsfonts}
\usepackage{amssymb}
\usepackage{graphicx}
\usepackage{placeins}
\usepackage{float}
\usepackage{algorithm}
\usepackage[noend]{algpseudocode}
\usepackage{natbib}
\usepackage{pstricks}
\usepackage{verbatim}
\usepackage{subfigure}
\usepackage{booktabs}
\usepackage{siunitx,booktabs,multirow, dcolumn}
\usepackage{setspace}

\newcommand{\ltsima} {$\; \buildrel < \over \sim \;$}
\newcommand{\gtsima} {$\; \buildrel > \over \sim \;$}
\newcommand{\lta} {\lower.5ex\hbox{\ltsima}}
\newcommand{\gta} {\lower.5ex\hbox{\gtsima}}

\newcommand{\tblspacer}{\vspace{.4em}}

\newcolumntype{d}[1]{D{.}{.}{#1} }

\def\f{\frac}
\def\nn{\nonumber}
\def\ln{\mathrm{ln}}
\def\exp{\mathrm{exp}}
\def\cov{\mathbf{C}}
\def\noise{\mathbf{N}}
\def\isotropic{\mathbf{T}}
\def\aniso{\bar{\mathbf{N}}}
\def\field{\mathbf{s}}
\def\data{\mathbf{d}}
\def\messenger{\mathbf{t}}
\def\bs{\boldsymbol}
\def\ie{i.e.}
\def\eg{e.g.}
\def\cf{cf.}
\def\lsim{\lesssim}
\def\gsim{\gtrsim}

\begin{document}

\title[Cosmological parameters from CFHTLenS weak lensing using Bayesian hierarchical inference]{Cosmological parameters, shear maps and power spectra from CFHTLenS using Bayesian hierarchical inference}

\author[J. Alsing, A. Heavens, A. Jaffe]
{\parbox{\textwidth}{Justin Alsing$^1$\thanks{e-mail:  j.alsing12@imperial.ac.uk}, 
Alan Heavens$^1$ and
Andrew H. Jaffe$^1$}\vspace{0.4cm}\\
\parbox{\textwidth}{$^1$ Imperial Centre for Inference and Cosmology, Department of Physics, Imperial College London, Blackett Laboratory, Prince Consort Road, London SW7 2AZ, UK}}
\date{Accepted ;  Received ; in original form }

\maketitle

\begin{abstract}
We apply two Bayesian hierarchical inference schemes to infer shear power spectra, shear maps and cosmological parameters from the CFHTLenS weak lensing survey --- the first application of this method to data. In the first approach, we sample the joint posterior distribution of the shear maps and power spectra by Gibbs sampling, with minimal model assumptions. In the second approach, we sample the joint posterior of the shear maps and cosmological parameters, providing a new, accurate and principled approach to cosmological parameter inference from cosmic shear data. As a first demonstration on data we perform a 2-bin tomographic analysis to constrain cosmological parameters and investigate the possibility of photometric redshift bias in the CFHTLenS data. Under the baseline $\Lambda$CDM model we constrain $S_8 = \sigma_8(\Omega_\mathrm{m}/0.3)^{0.5} = 0.67 ^{\scriptscriptstyle+ 0.03 }_{\scriptscriptstyle- 0.03 }$ $(68\%)$, consistent with previous CFHTLenS analyses but in tension with \emph{Planck} \citep{Planck2015XIII}. Adding neutrino mass as a free parameter we are able to constrain $\sum m_\nu < 4.6\mathrm{eV}$ (95\%) using CFHTLenS data alone. Including a linear redshift dependent photo-$z$ bias $\Delta z = p_2(z - p_1)$, we find $p_1=-0.25 ^{\scriptscriptstyle+ 0.53 }_{\scriptscriptstyle- 0.60 }$ and $p_2 =  -0.15 ^{\scriptscriptstyle+ 0.17 }_{\scriptscriptstyle- 0.15 }$, and tension with \emph{Planck} is only alleviated under very conservative prior assumptions. Neither the non-minimal neutrino mass or photo-$z$ bias models are significantly preferred by the CFHTLenS (2-bin tomography) data.
\end{abstract}
\section{Introduction}

Light from distant galaxies is continuously deflected by the gravitational potential fluctuations of large-scale structure on its way to us, resulting in a coherent distortion of observed galaxy images across the sky --- weak gravitational lensing. This weak lensing effect is a function of both the geometry of the Universe (through the distance-redshift relation) and the growth of potential fluctuations along the line-of-sight, making it a tremendously rich cosmological probe; the statistics of the weak lensing fields are sensitive to the initial conditions of the potential fluctuations, the relative abundance of baryonic and dark matter (through baryon acoustic oscillations), the linear and non-linear growth of structure, the mass and hierarchy of neutrinos (\eg, \citealp{Jimenez2010}), dark energy and gravity on large scales (see, \eg, \citealp{Weinberg2013} for a review). The goal of cosmic shear analyses is to extract cosmological inferences from the statistics of the observed weak lensing shear field --- the distortion of observed galaxy shapes measured across the sky and in redshift.

In \citet{Alsing2016} we developed a Bayesian hierarchical modelling (BHM) approach to infer the cosmic shear power spectrum (and thus cosmological parameters) from a catalogue of measured galaxy shapes and redshifts, building on previous work on cosmic microwave background (CMB) power spectrum inference \citep{Wandelt2004, Jewell2004, Eriksen2004, ODwyer2004, Chu2005, Larson2007, Eriksen2007} and large scale-structure analysis methods \citep{Jasche2010, Jasche2012, Jasche2013, Jasche2015}. The Bayesian hierarchical approach has a number of desirable features and advantages over traditional estimator-based methods: In contrast to frequentist estimators whose likelihoods need calibrating against large numbers of forward simulations (introducing assumptions and uncertainties that are often hard to propagate), the Bayesian approach explores the posterior distribution of the parameters of interest directly with clearly stated (and minimal) model assumptions, without the need for calibration. The Bayesian approach is exact and optimal, up to our ability to model the cosmic shear statistics (and systematics). Masks and complicated survey geometry are dealt with exactly and cleanly, in contrast to, \eg, pseudo-$C_\ell$ estimators that must carefully correct for mixing of $E$- and $B$-modes and physical (angular) scales arising from the mask inversion, which can be difficult in practice. The BHM approach can be readily extended to include models of non-Gaussian fields, exploiting more of the information-content of the weak lensing fields than is possible through $n$-point statistic estimators \citep{Jasche2013, Leclercq2015, Carron2012}. More generally, the BHM approach can also be extended to incorporate more of the weak lensing inference pipeline (\eg, shape measurement, PSF modelling etc), formally marginalising over nuisance parameters and systematics in a principled way and ultimately leading to more robust science at the end of the day (see \citet{Alsing2016, Schneider2015} for a discussion of the global hierarchical modelling approach to weak lensing).

The joint map-power spectrum inference approach proceeds in two distinct steps: In step one we sample the joint posterior of the shear map and power spectrum\footnote{We use ``power spectrum" as a shorthand for the full set of $EE$, $BB$ and $EB$ auto- and cross-power spectra for all tomographic bins.}, using, \eg, the Gibbs sampling approach described in \citet{Alsing2016}. In step two, we construct a smooth posterior density from the power spectrum samples and proceed to infer cosmological parameters by Markov Chain Monte Carlo (MCMC) sampling the power spectrum posterior as a function of the cosmological parameters (under our model of interest). The posterior distribution of the power spectrum is a valuable intermediate product; cosmological parameter inference can be performed for a large number of cosmological (and systematics) models directly from the power spectrum posterior \emph{a posteriori}, without loss of information and without having to re-analyse the entire data-set (since the initial power spectrum inference was independent of cosmology, assuming only statistical isotropy of the lensing fields). However, the need to estimate the continuous posterior density from a set of posterior samples may come with practical challenges; this \emph{density estimation} step may introduce uncertainties at some level, and accurate density estimation may be challenging for analyses with a large number of tomographic bins (for example, with $10$ tomographic bins we may need to estimate the joint distribution of the $55$ tomographic cross-power spectra from a set of MCMC samples --- a challenging density estimation task).

In this paper, we develop a second Bayesian approach to cosmic shear inference, whereby we jointly sample the shear maps and cosmological parameters, rather than the maps and power spectra. By going straight to cosmological parameters and bypassing the explicit power spectrum inference step, we circumvent the need to transform posterior samples into a continuous posterior density and hence avoid the prickly (high-dimensional) density estimation issues altogether. There are other advantages, too: by parametrising the power spectrum with a handful of cosmological parameters, the number of interesting parameters has been reduced from a few thousand power spectrum coefficients to typically $\lsim 10$ cosmological parameters -- this reduction in the parameter space will inevitably improve the sampling efficiency. Map-cosmology inference also extends more naturally to incorporate models for non-Gaussian shear where the power spectrum no longer fully specifies the lensing statistics. These benefits come at the cost of having to re-analyse the full shear maps for every cosmological model of interest, whereas the power spectrum posterior obtained in a cosmology-independent way represented a highly compressed intermediate product that could be used for (fast) \emph{a posteriori} cosmological parameter inference without the need to revisit the full data-set. However, we argue that nonetheless the map-cosmology inference scheme is a (comfortably) computationally feasible approach for current and future surveys.

In this paper we apply the map-power spectrum and map-cosmology sampling schemes to infer power spectra, shear maps and cosmological parameters from the Canada-France-Hawaii Telescope (CFHTLenS) weak lensing survey - the first application to data. The structure of the paper is as follows: In \S \ref{sec:data} we describe the CFHTLenS data and compression of the full galaxy catalogue into pixelized (noisy) tomographic shear maps. In \S \ref{sec:formalism} we review the tomographic weak lensing formalism, and in \S \ref{sec:bayes_schemes} we describe the map-power spectrum and map-cosmology Gibbs sampling schemes. In \S \ref{sec:implementation} we outline the cosmological and systematics models considered in this analysis. In \S \ref{sec:results} we demonstrate the Bayesian inference schemes on simulations before presenting the inferred $E$- and $B$- mode power spectra, tomographic shear maps and cosmological parameters from the CFHTLenS cosmic shear data. We discuss computation costs and prospects for future surveys in \S \ref{sec:cost} and conclude in \S \ref{sec:conclusions}.

\section{CFHTLenS data}
\label{sec:data}
CFHTLenS is a $154$ square degree optical imaging survey over four wide fields ($W1, W2, W3, W4$) in $ugriz$ bands \citep{Erben2013, Heymans2012}. The public catalogues\footnote{http://www.cfhtlens.org/astronomers/data-store} are a combination of data processing with \textsc{theli} \citep{Erben2013}, weak lensing measurements with \textsc{lensfit} \citep{Miller2013}, and photometric redshift (photo-$z$) posteriors using the Bayesian photo-$z$ code \textsc{bpz} \citep{Benitez2000, Hildebrandt2012}. We mask out CFHTLenS tiles that failed the systematics tests outlined in \citet{Heymans2012}, resulting in a removal of $25\%$ of the data. Stellar halos and image artefacts result in a further $\sim20\%$ of the remaining area being masked. We also restrict our analysis to the photo-$z$ range $0.5 < z_B < 1.3$, considered to be reliable by systematics tests outlined in \citet{Hildebrandt2013} and \citet{Benjamin2013}, where $z_B$ is the maximum posterior redshift provided by \textsc{bpz}.
\subsection{CFHTLenS Maps}
\label{sec:maps}
\begin{figure}
\begin{center}
\includegraphics[width = 8cm]{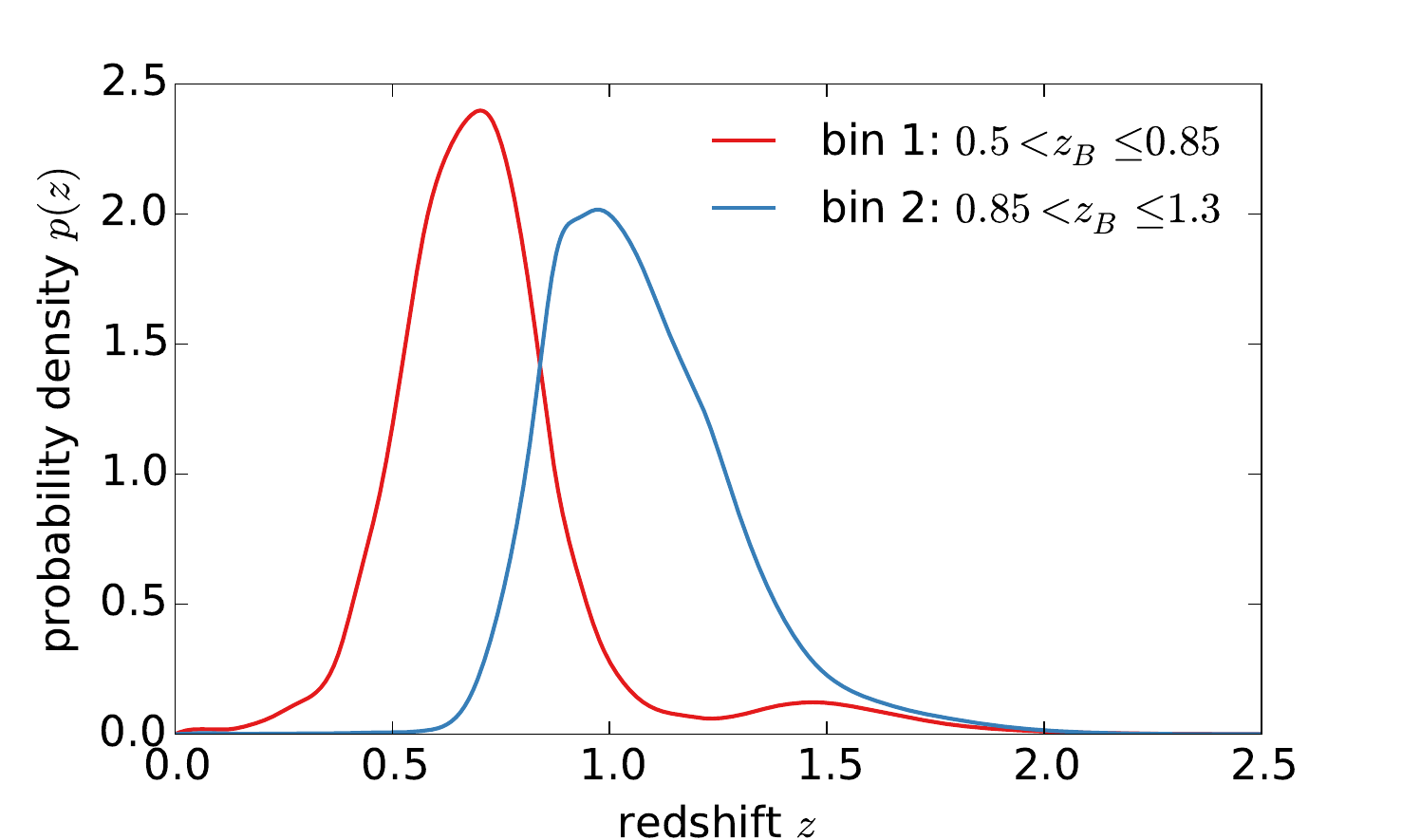}
\caption{Redshift distributions for tomographic bins $0.5 < z_B \leq 0.85$ and $0.85 < z_B \leq 1.3$, with sources separated by their best-fit \textsc{bpz} photometric redshifts $z_B$. The distributions $n_\alpha(z)$ are constructed by stacking the redshift posteriors for each individual source together (in each bin), weighted by their \textsc{lensfit} weights, see Eq. \eqref{pz}.}
\label{fig:cfhtlens_photo_z_dist}
\end{center}
\end{figure}

The resulting catalogue of $\num[group-separator={,}]{3099988}$ galaxies is processed into tomographic shear maps as follows: Following \citet{Benjamin2013} we divide the sources into two broad tomographic bins that are expected to be free from significant intrinsic alignment contamination: $0.5 < z_B \leq 0.85$ and $0.85 < z_B \leq 1.3$. The redshift distributions for each slice are estimated by stacking the individual source photo-$z$ posteriors $p(z)$ with their respective \textsc{lensfit} weights $w$ (shown in Figure \ref{fig:cfhtlens_photo_z_dist}),
\begin{align}
\label{pz}
n_\alpha(z) = \f{\sum_{g\in\mathrm{bin}\;\alpha}w_g p_g(z)}{\sum_{g\in\mathrm{bin}\;\alpha}w_g}.
\end{align}
Note that although use of stacked redshift posteriors for $n_\alpha(z)$ is commonplace in weak lensing analyses, this is not formally the correct statistical approach (see e.g., \citealp{Leistedt2016}). Later we will consider the possibility that the $n_\alpha(z)$ constructed from posterior-stacking may be biased (see \S \ref{sec:implementation} and \S \ref{sec:results}).

The four wide fields are pixelized into $175\times 175$, $113 \times 113$, $221 \times 221$ and $131\times 131$ square pixels respectively, with pixels of side length $\sigma_\mathrm{pix} = 3.82\;\mathrm{arcmin}$ in all cases. This pixelization scheme includes a $\sim 1$ degree border around the edge of each patch to reduce the impact of periodic boundary conditions (later assumed). 

\begin{figure*}
\includegraphics[width = 14cm]{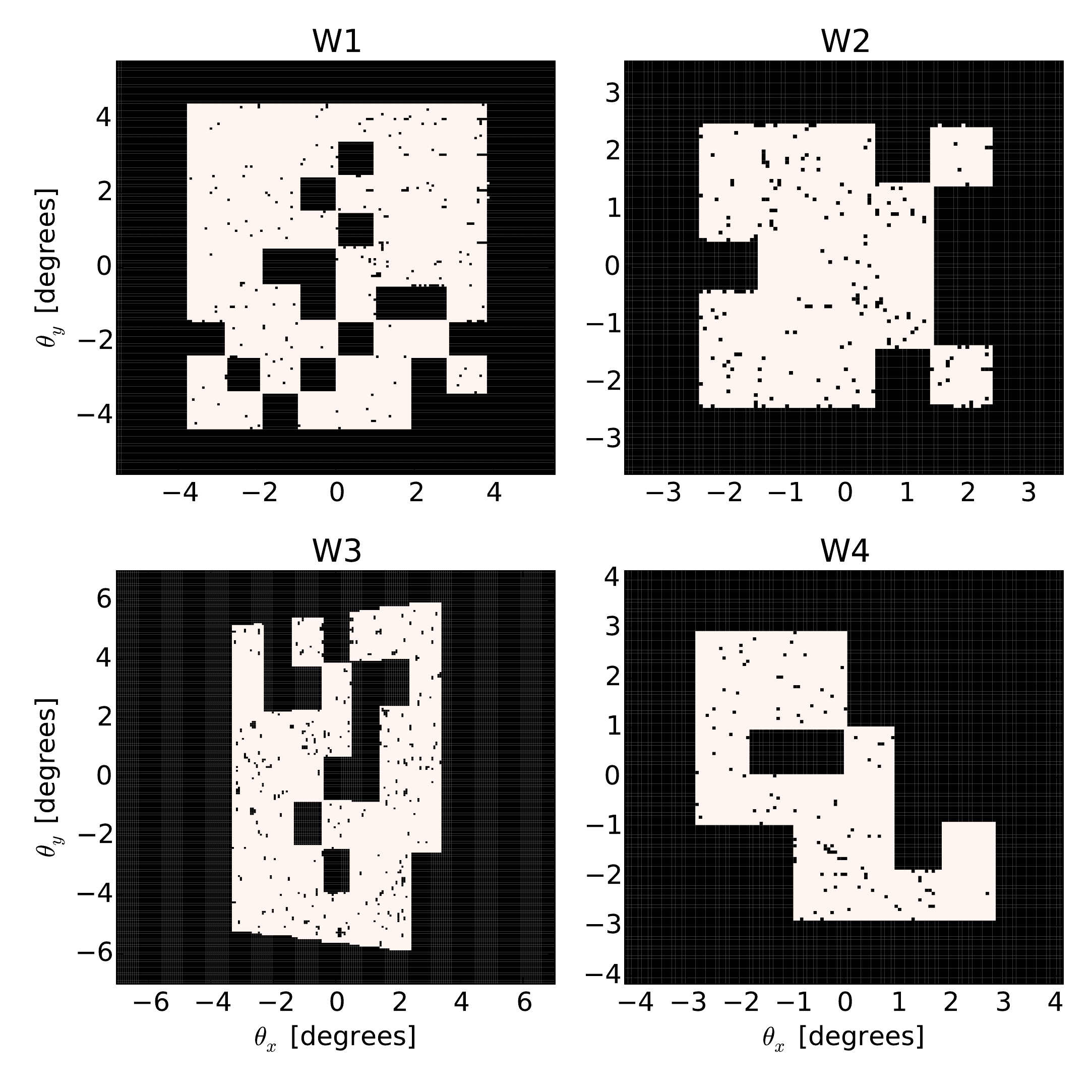}
\caption{The effective mask for the four CFHTLenS wide fields; black indicates masked regions (with no contributing sources). Each patch includes a $\sim1$ degree border to mitigate the impact of periodic boundary conditions (later assumed).}
\label{fig:cfhtlens_mask}
\end{figure*}

The angular coordinates of each source on the sky ($\mathrm{RA}$ $\alpha$ and $\mathrm{DEC}$ $\delta$) are converted to tangent-plane/flat-sky coordinates $(\theta_x, \theta_y)$ using a gnomonic projection: $\cos\theta_x = \cos^2(\pi/2-\delta) + \sin^2(\delta - \pi/2)\cos\alpha$ and $\theta_y = \delta$, where each patch is projected separately about its central coordinates. The estimated shear in each pixel $p$ is then given by
\begin{align}
\hat{\gamma}^\alpha_{i,p} = \f{\sum_{g\in p,\alpha} w_g(\epsilon_{i,g} - c_{i,g})}{(1+m)_p\sum_{g\in p,\alpha} w_g}
\end{align}
for each shear component $i\in\{1, 2\}$, and following \citet{Miller2013} we apply an additive bias correction $c_{i,g}$ on a source-by-source basis and a multiplicative bias correction
\begin{align}
(1 + m)_p = \sum_{g\in p,\alpha} w_g (1 + m_g)/\sum_{g\in p,\alpha}w_g
\end{align}
on a pixel-by-pixel basis. We define our data vector $\data$ organised as
\begin{align}
\data = (\hat\gamma_{1,p=1}^{(1)}, \hat\gamma_{2,p=1}^{(1)}, \hat\gamma_{1,p=1}^{(2)}, \hat\gamma_{2,p=1}^{(2)}, \hat\gamma_{1,p=2}^{(1)},\hat\gamma_{2,p=2}^{(1)}, \hat\gamma_{1,p=2}^{(2)}, \dots).
\end{align}
The intrinsic dispersion of galaxy ellipticities is taken to be $\sigma_\epsilon = 0.279$ per component, as estimated from the data and following previous CFHTLenS analyses \citep{Heymans2013, Kitching20143D, Benjamin2013, Kohlinger2016}, so the noise covariance in each pixel is given by
\begin{align}
N^\alpha_{i,p} =  \f{\sum_{g\in p,\alpha} w_g^2\sigma_\epsilon^2}{(\sum_{g\in p,\alpha} w_g)^2},
\end{align}
and the pixel-space noise covariance is organised as
\begin{align}
\mathrm{\mathbf{N}} = (N_{\scriptscriptstyle 1,p=1}^{(1)}, N_{\scriptscriptstyle 2,p=1}^{(1)}, N_{\scriptscriptstyle 1,p=1}^{(2)}, N_{\scriptscriptstyle 2,p=1}^{(2)}, N_{\scriptscriptstyle 1,p=2}^{(1)}, N_{\scriptscriptstyle 2,p=2}^{(1)}, N_{\scriptscriptstyle 1,p=2}^{(2)}, \dots).
\end{align} 
Masked pixels, \ie, pixels with no sources, are taken to have infinite noise. The effective CFHTLenS mask (regions containing no sources and hence infinite noise) is shown in Figure \ref{fig:cfhtlens_mask}.
\section{Tomographic shear formalism}
\label{sec:formalism}
In this section we briefly review the tomographic cosmic shear formalism, defining essential notation for describing the Bayesian inference schemes in \S \ref{sec:bayes_schemes}.

The expansion coefficients and two-point statistics of the complex shear fields in tomographic bins $\{\alpha\}$ (split into $E$- and $B$-mode components) in the small survey-area/flat-sky approximation are given by:
\begin{align}
&\gamma^{\mathrm{E}(\alpha)}_{\bs{\ell}} = \f{1}{2}\cdot\f{1}{2\pi}\int\left[\gamma^\alpha(\bs{\phi})\varphi_{\bs{\ell}}^*e^{-i\bs{\ell}\cdot\bs{\phi}} + \gamma^{*(\alpha)}(\bs{\phi})\varphi_{\bs{\ell}}e^{-i\bs{\ell}\cdot\bs{\phi}}\right]d\Omega,\nn \\
&\gamma^{\mathrm{B}(\alpha)}_{\bs{\ell}} = -\f{i}{2}\cdot\f{1}{2\pi}\int\left[\gamma^\alpha(\bs{\phi})\varphi_{\bs{\ell}}^*e^{-i\bs{\ell}\cdot\bs{\phi}} - \gamma^{*(\alpha)}(\bs{\phi})\varphi_{\bs{\ell}}e^{-i\bs{\ell}\cdot\bs{\phi}}\right]d\Omega,\nn \\
&\langle\gamma^{\mathrm{E}(\alpha)*}_{\bs{\ell}}\gamma^{\mathrm{E}(\beta)}_{\bs{\ell}'}\rangle = C^\mathrm{EE}_{\ell,\alpha\beta}\delta_{{\bs{\ell}}{\bs{\ell}}'}, \nn \\
&\langle\gamma^{\mathrm{E}(\alpha)*}_{\bs{\ell}}\gamma^{\mathrm{B}(\beta)}_{\bs{\ell}'}\rangle = C^\mathrm{EB}_{\ell,\alpha\beta}\delta_{{\bs{\ell}}{\bs{\ell}}'}, \nn \\
&\langle\gamma^{\mathrm{B}(\alpha)*}_{\bs{\ell}}\gamma^{\mathrm{B}(\beta)}_{\bs{\ell}'}\rangle = C^\mathrm{BB}_{\ell,\alpha\beta}\delta_{{\bs{\ell}}{\bs{\ell}}'},
\end{align}
where $\bs\ell = (\ell_x, \ell_y)$, the phase factor $\varphi_{\bs{\ell}} = - (\ell_x^2 - \ell_y^2 + 2i\ell_x\ell_y)/\ell^2$ and $C^\mathrm{EE}_{\ell,\alpha\beta}$, $C^\mathrm{EB}_{\ell,\alpha\beta}$ and $C^\mathrm{EE}_{\ell,\alpha\beta}$ are the $E$-mode, $B$-mode and cross $EB$ angular power spectra between tomographic bins $\alpha$ and $\beta$. Whilst cosmological models predict negligible $B$-modes and parity considerations require $C^\mathrm{EB}_{\ell,\alpha\beta} = 0$, systematic effects could give rise to non-zero $B$-modes and parity violating effects, so estimation of the $B$-mode power spectrum is useful as a test for residual systematics.

In the Limber approximation \citep{Limber1954}, the $E$-mode tomographic shear power spectra are given by \citep{Kaiser1992, Kaiser1998, Hu1999, Hu2002a, Takada2004},
\begin{align}
\label{limber_power}
C^\mathrm{EE}_{\ell,\alpha\beta} &= \int \f{d\chi}{\chi^2}\;w_\alpha(\chi)w_\beta(\chi)(1+z)^2P_\delta\left(\f{\ell}{\chi}; \chi\right)W_\ell^2,
\end{align}
where $\chi$ is comoving distance, $P_\delta(k; \chi)$ is the 3D matter power spectrum and we have assumed a spatially flat universe throughout. $W_\ell$ is the (azimuthally averaged) pixel window function
\begin{align}
W_\ell = \f{1}{N_\ell}\sum_{|\bs{\ell}| = \ell}\f{\sin(\sigma_\mathrm{pix}\ell_x/2)}{\sigma_\mathrm{pix}\ell_x/2}\f{\sin(\sigma_\mathrm{pix}\ell_y/)2}{\sigma_\mathrm{pix}\ell_y/2},
\end{align}
where $N_\ell$ $(\ell_x, \ell_y)$-modes contribute to mode $\ell$ and the average is performed over the modes that appear in the analysis (from the pixelized patches in Fourier space). The lensing weight functions $w_\alpha(\chi)$ are given by
\begin{align}
w_\alpha(\chi)=\f{3\Omega_\mathrm{m}H_0^2}{2}\chi\int_{\chi}^{\chi_\mathrm{H}} d\chi'\;n_\alpha(\chi')\f{\chi'-\chi}{\chi'}
\end{align}
where $n_\alpha(\chi)d\chi = p_\alpha(z)dz$ is the redshift distribution for galaxies in redshift bin $\alpha$ (normalized to unity over the bin), i.e., Eq. \eqref{pz}.

We define the shear field vector $\field$ as the collection of tomographic shear maps $\{\gamma^\alpha(\theta,\phi)\}$, organised as
\begin{align}
&\field = \left(\field_{\bs{\ell}_1}, \field_{\bs{\ell}_2}, \field_{\bs{\ell}_3}, \dots\field_{\bs{\ell}_i}\dots\right), \nn \\
&\field_{\bs{\ell}} = \left(\gamma^{\mathrm{E}(1)}_{\bs{\ell}}, \gamma^{\mathrm{E}(2)}_{\bs{\ell}}, \dots, \gamma^{\mathrm{E}(n_\mathrm{bins})}_{\bs{\ell}}, \gamma^{\mathrm{B}(1)}_{\bs{\ell}}, \gamma^{\mathrm{B}(2)}_{\bs{\ell}}, \dots \gamma^{\mathrm{B}(n_\mathrm{bins})}_{\bs{\ell}}\right).
\end{align}
The full covariance matrix $\cov$ of the field $\field$ will be block-diagonal, with each $\bs\ell$-mode contributing one block,
\begin{align}
&\langle\field\field^\dagger\rangle = \cov = \mathrm{diag}\left(\cov_{\bs{\ell}_1}, \cov_{\bs{\ell}_2}, \cov_{\bs{\ell}_3}, \cov_{\bs{\ell}_4}\dots \cov_{\bs{\ell}_i}\dots\right),
\end{align}
where
\begin{align}
\label{flat_sky_covariance_matrix}
&\cov_{\ell_x \ell_y} = \left( \begin{array}{ccccccc}
C^{\mathrm{EE}}_{\ell, 11} & C^{\mathrm{EE}}_{\ell, 12} & \dots & C^{\mathrm{EB}}_{\ell, 11} & C^{\mathrm{EB}}_{\ell, 12} & \dots \\
C^{\mathrm{EE}}_{\ell, 21} & C^{\mathrm{EE}}_{\ell, 22} & \dots &C^{\mathrm{EB}}_{\ell, 21} & C^{\mathrm{EB}}_{\ell, 22} & \dots \\
\vdots & \vdots &  \ddots & \vdots& \vdots & \ddots \\
C^{\mathrm{BE}}_{\ell, 11} & C^{\mathrm{BE}}_{\ell, 12} & \dots & C^{\mathrm{EB}}_{\ell, 11} & C^{\mathrm{BB}}_{\ell, 12} & \dots \\
C^{\mathrm{BE}}_{\ell, 21} & C^{\mathrm{BE}}_{\ell, 22} & \dots &C^{\mathrm{BB}}_{\ell, 21} & C^{\mathrm{BB}}_{\ell, 22} & \dots \\
\vdots & \vdots &  \ddots & \vdots& \vdots & \ddots \end{array} \right).
\end{align}
\section{Bayesian inference schemes}
\label{sec:bayes_schemes}
In this section we describe two Bayesian hierarchical inference approaches for extracting cosmological information from weak lensing surveys. In \S \ref{sec:map_power} we review the map-power spectrum inference scheme presented in \citet{Alsing2016}, and in \S \ref{sec:messenger_hierarchical_model} we develop a new hierarchical inference scheme for sampling the joint posterior of the map and cosmological parameters, building on an approach developed by \citet{Racine2015} and \citet{Jewell2009}. For a detailed pedagogical account of the map-power spectrum sampling approach (including a detailed discussion of the messenger field sophistication), see \citet{Alsing2016}.
\subsection{Hierarchical map-power spectrum inference}
\label{sec:map_power}
\begin{figure}
\begin{center}
\begin{subfigure}
\centering
\includegraphics[width=3cm]{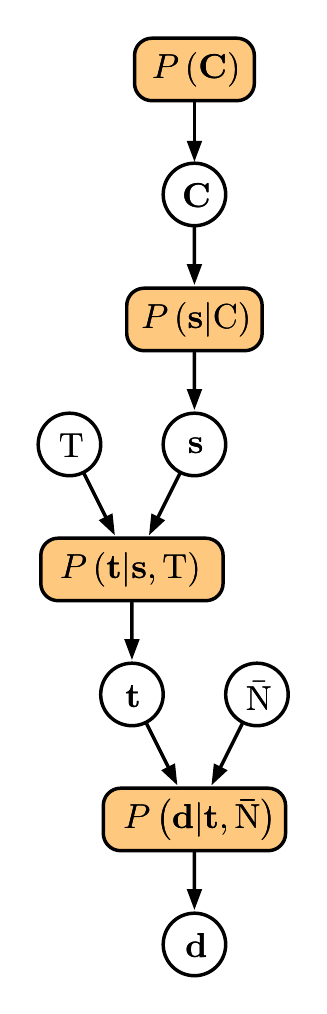}
\end{subfigure}
\begin{subfigure}
\centering
\includegraphics[width=3.2cm]{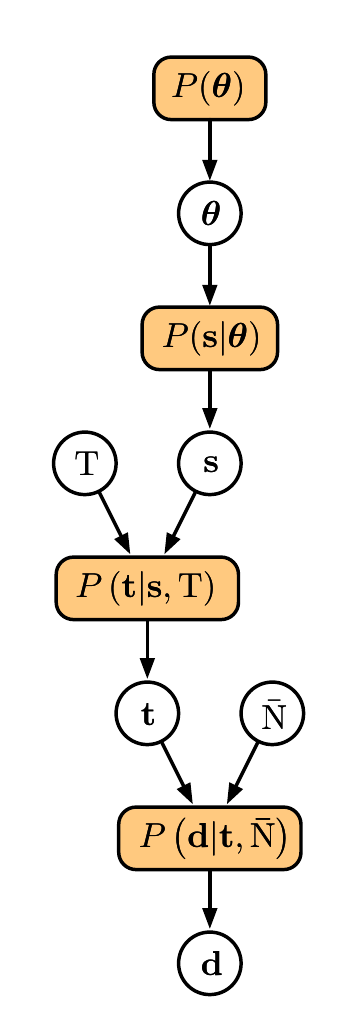}
\end{subfigure}
\end{center}
\caption{Left: Hierarchical forward model for shear map-power spectrum inference (with the messenger field sophistication): the shear power spectrum $\cov$ is drawn from some prior distribution, a realization of the shear field $\field$ is then generated given the power spectrum, isotropic noise with covariance $\isotropic$ is added to give a realization of the messenger field $\messenger$ and finally an anisotropic noise component is added with covariance $\aniso$ to realize the noisy shear maps (i.e., data) $\data$. Right: Hierarchical forward model for shear map-cosmological parameter inference, where the power spectrum $\cov$ has been replaced by the cosmological parameters $\bs\theta$.}
\label{fig:messenger_hierarchical_model}
\end{figure}
The generative forward model for the cosmic shear data (via the signal covariance $\cov$, shear field $\field$ and messenger field $\messenger$) is summarised in Fig. \ref{fig:messenger_hierarchical_model} and can be understood as follows: the signal covariance $\cov$ generates tomographic shear maps $\field$, the shear maps plus isotropic noise with covariance $\isotropic$ generate a realisation of the messenger field $\messenger$, and finally the messenger field plus an additional anisotropic noise component with covariance $\aniso$ generates a realisation of the data $\data$. The full noise covariance is given by the sum of the isotropic and anisotropic parts $\mathbf{N} = \isotropic + \aniso$ where, following \cite{Elsner2012}, we define $\isotropic = \tau\mathbf{I}$ with $\tau = \mathrm{min}\left[\mathrm{diag}(\mathbf{N})\right]$ (\ie, the largest isotropic noise component that can be extracted from $\mathbf{N}$)\footnote{More generally if the noise covariance is not diagonal, one should take $\tau$ to be the largest value that leaves $\aniso = \mathbf{N} - \isotropic$ positive definite.}. The introduction of the auxiliary messenger field separates the signal covariance $\cov$ from the anisotropic noise covariance $\aniso$ in the hierarchy, connecting them only via $\isotropic\propto\mathbf{I}$ which is diagonal in any basis; this is the essential function of the messenger field and enables us to perform all necessary matrix inversions in bases where the matrices are sparse \citep{Jasche2015, Alsing2016}, realising a dramatic computational improvement over earlier Gibbs sampling approaches \citep{Wandelt2004, Jewell2004, Eriksen2004, ODwyer2004, Chu2005, Larson2007, Eriksen2007}.

In this study, we will assume that both the shear field $\field$ and the noise are Gaussian distributed. Under these assumptions, the probability densities appearing in the graph (Fig. \ref{fig:messenger_hierarchical_model}) are given by
\begin{align}
&P(\data|\messenger, \aniso) =\frac{1}{\sqrt{(2\pi)^N|\aniso|}}e^{-\frac{1}{2}(\mathbf{d}-\messenger )^\dagger \aniso^{-1}(\mathbf{d}-\messenger )}, \nn \\
&P(\field |\cov) = \frac{1}{\sqrt{(2\pi)^N|\cov|}}e^{-\frac{1}{2}\field^\dagger \cov^{-1}\field }, \nn \\
&P(\messenger|\field, \isotropic) = \frac{1}{\sqrt{(2\pi)^N|\isotropic|}}e^{-\frac{1}{2}(\messenger-\field)^\dagger \isotropic^{-1}(\messenger -\field)},
\end{align}
where $N = 2\times n_{bins}\times n_{pix}$ is the length of the vectors $\data$, $\field$ and $\messenger$ for $n_{bins}$ tomographic shear maps each containing $n_{pix}$ pixels (and the factor of 2 is due to $E$- and $B$-mode degrees-of-freedom). Note that marginalising the posterior $P(\cov, \field, \messenger|\data)$ over the field $\field$ and messenger field $\messenger$ recovers the posterior distribution signal covariance (power spectrum) $P(\cov|\data)$ as desired.

In order to Gibbs sample from the posterior $P(\cov, \field, \messenger|\data)$ we must iteratively draw samples from $\cov$, $\field$ and $\messenger$ conditional on all other parameters, \ie, 
\begin{align}
&\cov^{i+1} \leftarrow P(\cov|\field^i), \nn \\
&\field^{i+1} \leftarrow P(\field|\messenger^i, \cov^{i+1}, \isotropic), \nn \\
&\messenger^{i+1} \leftarrow P(\messenger|\field^{i+1}, \data, \aniso, \isotropic).
\end{align}
where the conditional distributions are given by:
\begin{align}
\label{messenger_conditionals}
P(\field|\cov, \isotropic, \messenger) &\propto P(\messenger|\field, \isotropic)P(\field|\cov) \nn \\
&= \frac{1}{\sqrt{(2\pi)^N|\mathbf{Q}_\field|}}e^{-\frac{1}{2}(\field - \bs{\mu}_\field )^\dagger \mathbf{Q}_\field^{-1}(\field - \bs{\mu}_\field )}, \nn \\
P(\messenger|\field, \isotropic, \aniso, \data) &\propto P(\data|\messenger, \aniso)P(\messenger|\isotropic, \field) \nn \\
&= \frac{1}{\sqrt{(2\pi)^N|\mathbf{Q}_\messenger|}}e^{-\frac{1}{2}(\messenger - \bs{\mu}_\messenger)^\dagger \mathbf{Q}_\messenger^{-1}(\messenger - \bs{\mu}_\messenger)}, \nn \\
P(\cov_\ell|\mathbf{s}) & \propto P(\field | \cov) P(\cov)
=\mathcal{W}^{-1}(\bs{\Gamma}_\ell, \nu_\ell).
\end{align}
$\mathcal{W}^{-1}(\bs{\Gamma}_\ell, \nu_\ell)$ denotes the inverse Wishart distribution with support $\bs{\Gamma}_\ell = \sum_{| \bs\ell |=\ell}  \field_{\bs\ell}\field_{\bs\ell}^\dagger$ and $\nu_\ell = (\sum_{| \bs\ell |=\ell}) - p - 1$ degrees-of-freedom for a $p\times p$ covariance matrix, and we have assumed a uniform prior over the signal covariance $P(\cov) = \mathrm{constant}$. The (conditional) shear field and messenger field means and covariances are given by,
\begin{align}
\label{mean_covars}
\bs{\mu}_\field &= (\cov^{-1} + \isotropic^{-1})^{-1}\isotropic^{-1}\messenger, \nn \\
\mathbf{Q}_\field & = (\cov^{-1} + \isotropic^{-1})^{-1}, \nn \\
\bs{\mu}_\messenger &= (\isotropic^{-1} + \aniso^{-1})^{-1}\isotropic^{-1}\field + (\isotropic^{-1} + \aniso^{-1})^{-1}\aniso^{-1}\data, \nn \\
\mathbf{Q}_\messenger & = (\isotropic^{-1} + \aniso^{-1})^{-1}.
\end{align}

Drawing samples from the shear field, messenger field and power spectrum conditionals in Eq. \eqref{messenger_conditionals} is then straightforward: Constrained shear map realisations drawn as Gaussian random variates, with mean $\bs{\mu}_\field$ and covariance $\mathbf{Q}_\field$ given in Eq. \eqref{mean_covars}. Similarly, constrained messenger field realisations are drawn as Gaussian random variates with mean $\bs{\mu}_\messenger$ and covariance $\mathbf{Q}_\messenger$. Power spectrum (signal covariance) samples are drawn as inverse Wishart random variates $\cov_\ell \leftarrow \mathcal{W}^{-1}(\bs{\Gamma}_\ell, \nu_\ell)$, which can be generated as follows:
\begin{enumerate}
\item Generate $\nu_\ell$ Gaussian random vectors $\mathbf{x}_i\leftarrow\mathcal{N}(\mathbf{0}, \boldsymbol{\Gamma}_\ell^{-1})$.
\item Construct the sum of outer products of the vectors $\{\mathbf{x}_i\}$, \ie, $\mathbf{X} = \sum_{i=1}^\nu \mathbf{x}_i\mathbf{x}_i^\mathrm{T}$.
\item Take the inverse of $\mathbf{X}$, then $\mathbf{X}^{-1}\sim\mathcal{W}^{-1}(\boldsymbol{\Gamma}_\ell, \nu_\ell)$ as required.
\end{enumerate}
We will apply the map-power spectrum sampling scheme to infer tomographic shear maps and power spectra from CFHTLenS data in \S \ref{sec:results}. In the following section we discuss how to perform lossless cosmological parameter inference from the power spectrum posterior.
\subsubsection{Cosmological parameter inference from the power spectrum posterior}
Cosmological models provide a deterministic mapping between cosmological parameters and the shear power spectrum, \ie, for a given set of cosmological parameters $\bs\theta$ (under a given model $\mathcal{M}$) we can compute the set of power spectrum coefficients $\cov_\ell \equiv \cov_\ell(\bs\theta, \mathcal{M})$. This means that if we have access to the posterior $P(\cov_\ell | \data)$, we can sample the posterior distribution of cosmological parameters without loss of information, by drawing samples from
\begin{align}
\label{cosmo_distr}
\bs\theta \leftarrow P(\cov_\ell(\bs\theta) | \data)P(\bs\theta)/P(\cov_\ell(\bs\theta)),
\end{align}
where notably the prior on the signal covariance has been effectively replaced by a prior over the cosmological parameters, so our initial choice of prior on $\cov$ becomes irrelevant. Hence, in order to perform cosmological parameter inference we need access to the full smooth posterior density $P(\cov_\ell | \data)$. The Gibbs map-power spectrum sampling approach described above generates a set of samples from the power spectrum posterior -- in order to sample from Eq. \eqref{cosmo_distr} we must hence estimate the density $P(\cov_\ell | \data)$ from the MCMC samples. There are a number of approaches one could take to this density estimation task -- kernel density estimators and mixture models provide fast and flexible schemes for estimating probability densities from a set of samples \citep{Silverman1986}, or alternatively the more specialized Blackwell-Rao estimator (tailored to Gibbs sampling output) has been shown to be effective in the context of CMB and other Gibbs samplers \citep{Gelfand1990, Chu2005}.

Whilst this density estimation step is conceptually simple, there are situations where it may be technically challenging and introduce uncertainties. For example, in an analysis with a large number of tomographic bins we may be faced with estimating a high-dimensional density $P(\cov_\ell | \data)$. For example, a 10-bin analysis has $55$ tomographic $E$-mode (cross) power spectrum coefficients per $\ell$ mode, which are likely to be correlated with one another. Even if we could treat each $\ell$-mode independently, we are faced with estimating a $55$-dimensional probability density from a set of samples, and if $\ell$-mode correlations in the posterior are non-negligible this number will increase even further. Whilst this is unlikely to be an insurmountable problem (particularly in the fortuitous case where the posteriors are close to Gaussian), we expect it to present some technical challenge and introduce uncertainties at some level. In the spirit of the Bayesian approach, making clearly stated assumptions and minimal approximations, we would like an alternative method that is free from this density estimation step. In \S \ref{sec:messenger_hierarchical_model} we develop an alternative (exact) Bayesian sampling approach that bypasses the density estimation step completely.

Note that frequentist power spectrum estimator methods suffer from a similar (and often harder) problem to the density estimation step in the Bayesian approach. Estimator methods need to reconstruct the sampling distribution (likelihood) of the chosen estimator $\hat{\cov}$, \ie, the distribution of the estimator given some true power spectrum $P(\hat{\cov} | \cov)$. This is typically done by generating a large number of samples of the estimator through forward simulations, and estimating the sampling distribution from those samples (analogous to the density-estimation problem described above). Reconstructing sampling distributions from forward simulations is generally harder than reconstructing posterior densities from MCMC samples; formally, the sampling distribution $P(\hat{\cov} | \cov)$ must be known for all plausible values of the true power spectrum $\cov$, whereas the estimating the posterior distribution represents a single density-estimation task from the set of MCMC samples (with the data fixed). As a result one usually resorts to further approximation schemes for constructing estimator likelihoods or covariances (see \citealp{Efstathiou2004} for a discussion).

For non-sampling based approaches for approximating the full likelihood function for the power spectrum (in the context of CMB power spectrum inference), see \citet{Bond1998, Bond2000, Hamimeche2008}.
%
%
%
In this section we develop a new Bayesian inference scheme whereby we explore the joint posterior of the shear maps and cosmological parameters, rather than the shear maps and power spectra. By going straight to cosmological parameters and bypassing the explicit power spectrum inference step, we circumvent the need to transform posterior samples into a continuous posterior density avoiding the (potentially challenging) density estimation step altogether. The map-cosmology sampling approach has some additional advantages: by parametrising the power spectrum by a handful of cosmological parameters, the number of interesting parameters has been reduced from thousands of power spectrum coefficients to typically $\lsim 10$ cosmological parameters --- this dramatic shrinking of the parameter space will inevitably improve the sampling efficiency. The map-cosmology inference also extends very naturally to models for non-Gaussian shear where the power spectrum no longer fully specifies the lensing statistics, allowing us to ultimately exploit more of the information content of the lensing fields (given a model for the non-Gaussian shear statistics). These benefits come at a cost; whereas the cosmology-independent power spectrum posterior represented a highly compressed intermediate product that could be used for (fast) \emph{a posteriori} cosmological parameter inference, the map-cosmology sampler assumes a cosmological model from the beginning and hence needs to be run on the full data-set for each model of interest.

The hierarchical forward model for joint map-cosmology inference is shown in Fig. \ref{fig:cosmology_hierarchical_model} and is understood as follows: the cosmological parameters $\bs\theta$ (drawn from some prior) specify a power spectrum $\cov(\bs\theta)$ from which a realization of the shear field $\field$ is generated, isotropic noise with covariance $\isotropic$ is added to give a realization of the messenger field $\messenger$ and finally an anisotropic noise component is added with covariance $\aniso$ to realize a noisy shear map $\data$. The parameters $\bs{\theta}$, $\field$ and $\messenger$ can be sampled by Gibbs sampling in the usual way,
\begin{align}
&\bs\theta^{i+1} \leftarrow P(\bs\theta | \field^i) \nn \\
&\field^{i+1} \leftarrow P(\field | \cov(\bs\theta^{i+1}),\messenger) \nn \\
&\messenger^{i+1} \leftarrow P(\messenger | \field^{i+1}, \data),
\end{align}
where the shear and messenger conditionals are identical to Eq. \eqref{messenger_conditionals}. Again making the assumption of Gaussian lensing fields, the cosmological parameters only enter via the power spectrum $\cov\equiv\cov(\bs\theta)$ and the cosmological parameter conditional is given by
\begin{align}
\label{cosmological_conditional}
P(\bs\theta | \field) &\propto P(\field | \cov(\bs\theta) ) P(\bs\theta)  \nn \\
&= P(\bs\theta)\times\prod_{\bs\ell} \f{1}{\sqrt{(2\pi)^p |\cov_{\ell}|}}\exp\left[-\f{1}{2}\field_{\bs\ell}^\dagger\cov^{-1}_\ell \field_{\bs\ell}\right],
\end{align}
where the sub-covariance matrices $\cov_\ell$ have size $p\times p$. Sampling from the shear and messenger field conditionals is straightforward (as we saw in \S \ref{sec:map_power}). Drawing samples from the cosmological parameters is more difficult; due to the non-linear mapping between the cosmological parameters and the power spectrum, the conditional $P(\bs\theta | \field)$ is not a simple distribution and cannot be sampled straightforwardly (in contrast to drawing Gaussian and inverse-Wishart variates in \S \ref{sec:map_power}). Instead, we must resort to an alternative sampling scheme, such as introducing a Metropolis-Hastings step for the $\bs\theta$ parameters. Nonetheless, sampling from the joint posterior $P(\bs\theta,\field,\messenger | \data)$ reduces to realising Gaussian random fields for $\field$ and $\messenger$ according to Eq. \eqref{messenger_conditionals} and a simple Metropolis-Hastings (MH) step for the cosmological parameters $\bs\theta$ (which is usually a small number of parameters compared to $\field$ and $\messenger$).

In the map-cosmology inference scheme described above we have assumed that the signal covariance $\cov$ is fully parametrised by cosmological parameters $\bs\theta$. However, (to leading order) cosmological weak lensing predicts $E$-mode power only and as such we have implicitly neglected $B$-modes (effectively assuming zero $B$-mode power). Including $B$-modes in the map-cosmology inference scheme can be easily achieved by either adding an additional inverse-Wishart sampling step for the $B$-mode power spectrum (\cf, Eq. \eqref{messenger_conditionals}), or including a model for the $B$-mode power and extending $\bs\theta$ to include the parameters of the $B$-mode model alongside the cosmological parameters. In the implementation of the map-cosmology inference scheme used in this paper, we consider $E$-modes only (where $B$-modes are rather constrained using the map-power spectrum inference scheme, \cf, \S \ref{sec:map_power}).
\subsubsection*{Efficient sampling in the low S/N regime}
The Gibbs sampling approach described above will jointly sample the shear map, messenger field and cosmological parameters. However, in the low signal-to-noise limit Gibbs sampling is expected to be inefficient; when sampling the cosmological parameters conditioned on the map $\bs\theta \leftarrow P(\bs\theta | \field)$, the step size will be determined by the cosmic variance, whilst the full posterior density is determined by the cosmic variance plus the noise variance. In the limit where the noise variance is totally dominant over the cosmic variance, a very large number of samples will be required to explore the full width of the posterior distribution. \citet{Jewell2009} made the first concerted effort to formulate a new sampling algorithm that overcomes the poor efficiency of the Gibbs sampler at low S/N (for map-power spectrum sampling) and recently \citet{Racine2015} developed a map-cosmology sampling scheme that is efficient across the full range of S/N. Here we will adopt the \citet{Racine2015} approach, extending their work to include a messenger field for efficient sampling in the presence of masks.

The \citet{Racine2015} approach replaces the simple Metropolis-Hastings $\bs\theta \leftarrow P(\bs\theta | \field)$ step with a joint map-cosmological parameter move that proceeds as follows: propose a new cosmology $\bs\theta^i$ from some proposal density $q(\bs\theta | \bs\theta^{i-1})$ and compute the updated power spectrum $\cov^i(\bs\theta^i)$. Then we construct a `rescaled' shear map, according to
\begin{align}
\label{rescaling}
\field^i = \hat{\field}^{i} + \cov_i^{1/2}\cov_{i-1}^{-1/2}(\field^{i-1} - \hat{\field}^{i-1})
\end{align}
where $\hat{\field}^{i}$ is the expectation value of the map-conditional $P(\field^{i} | \cov^{i}, \messenger^{i-1})$, \ie, $\hat{\field}^{i} = (\cov_{i}^{-1} + \isotropic^{-1})^{-1}\isotropic^{-1}\messenger^{i-1}$. Finally, we accept/reject the proposed $\bs\theta^i$ and rescaled map $\field^i$ with acceptance probability 
\begin{align}
\label{acceptance1}
A = \mathrm{min}\left[1, \f{\pi(\bs\theta^i)q(\bs\theta^{i-1}|\bs\theta^i)P(\bs\theta^i)}{\pi(\bs\theta^{i-1})q(\bs\theta^i |\bs\theta^{i-1})P(\bs\theta^{i-1})}\right]
\end{align}
where $P(\bs\theta)$ is the prior on the cosmological parameters, $q(\bs\theta^i |\bs\theta^{i-1})$ is the proposal density and $\pi(\bs\theta)$ is given by
\begin{align}
\label{acceptance2}
\pi(\bs\theta) = \exp\left[-\f{1}{2}(\messenger - \hat{\field})^\dagger \isotropic^{-1}\right.&(\messenger - \hat{\field})- \f{1}{2}\hat{\field}^\dagger\cov^{-1}\hat{\field} \nn \\
&\left. -\f{1}{2}(\field - \hat{\field})^\dagger\isotropic^{-1}(\field - \hat{\field})\right].
\end{align} 
For a detailed derivation of the acceptance ratio and a discussion of deterministic `rescaling' moves in the context of MCMC (including proof of detailed balance and irreducibility), see \citet{Racine2015} and \citet{Jewell2009}. The full sampling algorithm for $P(\bs\theta,\field,\messenger | \data)$ can then be summarized as repeated a sequence of three steps:
\begin{align*}
&\messenger^{i+1} \leftarrow P(\messenger | \field^i, \data), \nn \\
&\field^{i+1/2} \leftarrow P(\field | \messenger^{i+1}, \cov(\bs\theta^i)) \nn \\
&\{\bs\theta^{i+1},\field^{i+1} \}\leftarrow \mathcal{A}(\bs\theta, \field |  \field^{i+1/2}, \bs\theta^i ),
\end{align*}
where $\mathcal{A}(\bs\theta', \field' |  \field, \bs\theta )$ denotes the joint map-cosmology MH sampling move described above and we have introduced a ``half-step'' in the shear map sampling, since two map samples are generated per full Gibbs cycle.
\begin{figure}
\begin{center}
\includegraphics[width=8cm]{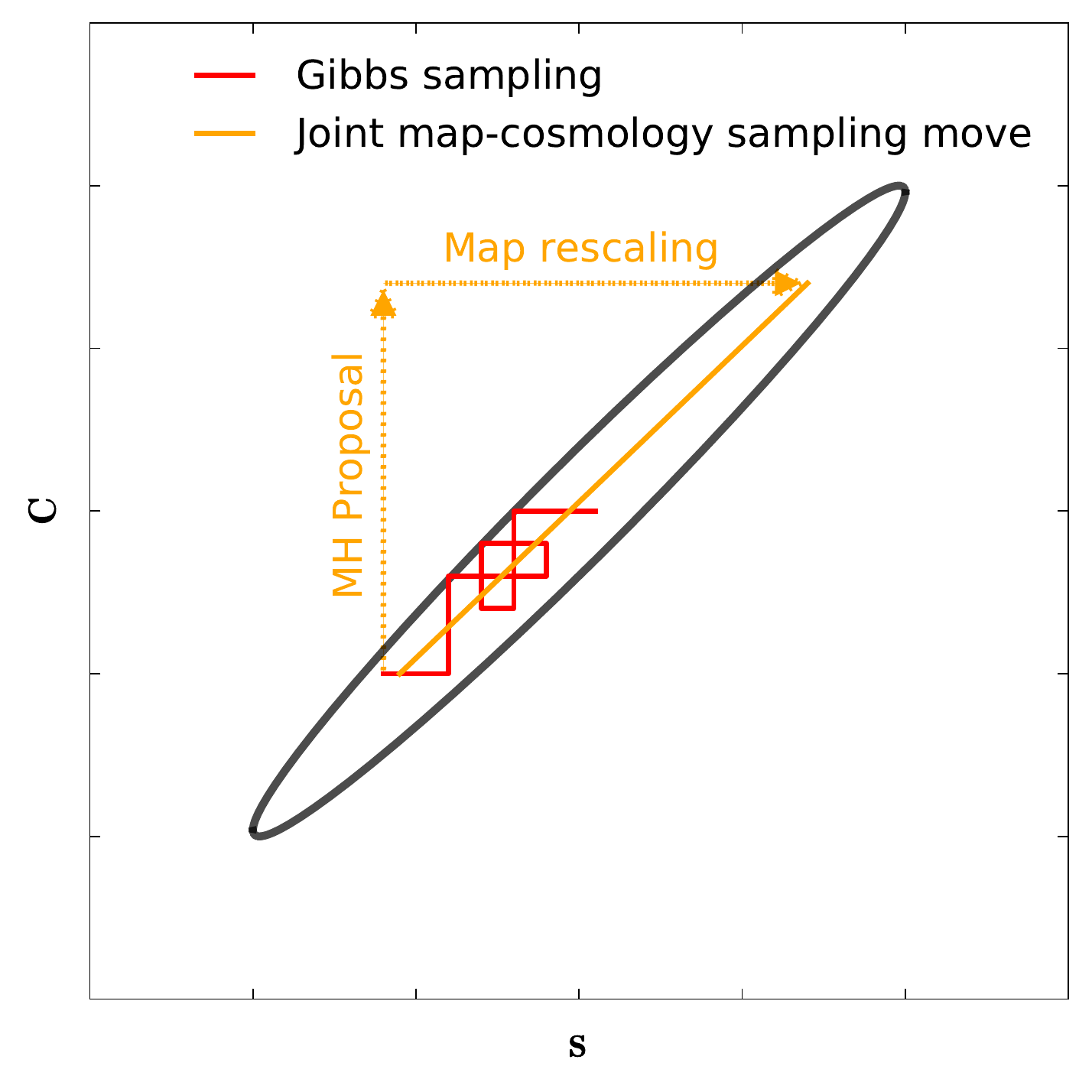}
\end{center}
\caption{Schematic demonstration of the joint map-cosmological parameter move introduced by \citet{Racine2015} for efficient sampling in the low S/N regime. In this regime, the width of the joint posterior at fixed $\field$ is representative of the cosmic variance, whilst the full width of the marginal posterior for $\cov$ is determined by the sum of the cosmic and noise variances. Gibbs sampling (red) is inefficient at exploring the full posterior, since it is constrained to move along the $\cov$ and $\field$ directions and can hence only take small steps comparable to the cosmic variance. In contrast, the joint sampling step proposes a new set of cosmological parameters (and hence $\cov$), and then rescales the map to bring the $\{\field,\cov\}$-pair back into a region of reasonable posterior density, where it will stand a good chance of being accepted by Metropolis-Hastings acceptance/rejection.}
\label{fig:racine}
\end{figure}

The joint map-cosmology move is designed to be efficient across the full range of S/N. The efficiency of the joint sampling move can be understood as follows (summarised schematically in Fig. \ref{fig:racine}): In the usual Gibbs sampling approach, when drawing cosmological parameters conditioned on the map, the step size will be limited by the cosmic variance. Similarly, constrained map sampling steps too can only move within the cosmic variance; whilst the full joint posterior is characterised by the cosmic variance plus noise, at any given Gibbs step the map and cosmological parameters must be consistent with each other to within the cosmic variance. In contrast, the joint map-cosmology sampling step proposes a new set of cosmological parameters (and hence $\cov$), and then \emph{re-scales} the map to bring the $\{\field,\cov\}$-pair back into a region of reasonable posterior density, where it will stand a good chance of being accepted by Metropolis-Hastings acceptance/rejection (see Fig. \ref{fig:racine}). As such, the joint map-cosmology move is able to make large steps and explore the full posterior width efficiently. In the high signal-to-noise regime, the map-rescaling of Eq. \eqref{rescaling} has negligible effect, and the joint map-cosmology sampling move reduces to a standard Metropolis-Hastings step in $\bs\theta$ (which we expect to be efficient in the high S/N regime where the posterior width is dominated by the cosmic variance). For a detailed pedagogical discussion of the joint map-cosmology sampling move, see \citet{Racine2015} and \citet{Jewell2009}.

Note that whilst we have applied the \citet{Racine2015} approach to map-cosmology inference, one could apply the same machinery for efficient map-power spectrum sampling by employing a joint map-power spectrum MH sampling move. We leave implementation of this approach to map-power spectrum sampling to future work, but highlight that it promises significant gains in sampling efficiency over the Gibbs sampling approach of \S \ref{sec:map_power} and should be pursued.
\begin{figure*}
\includegraphics[width = 17.5cm]{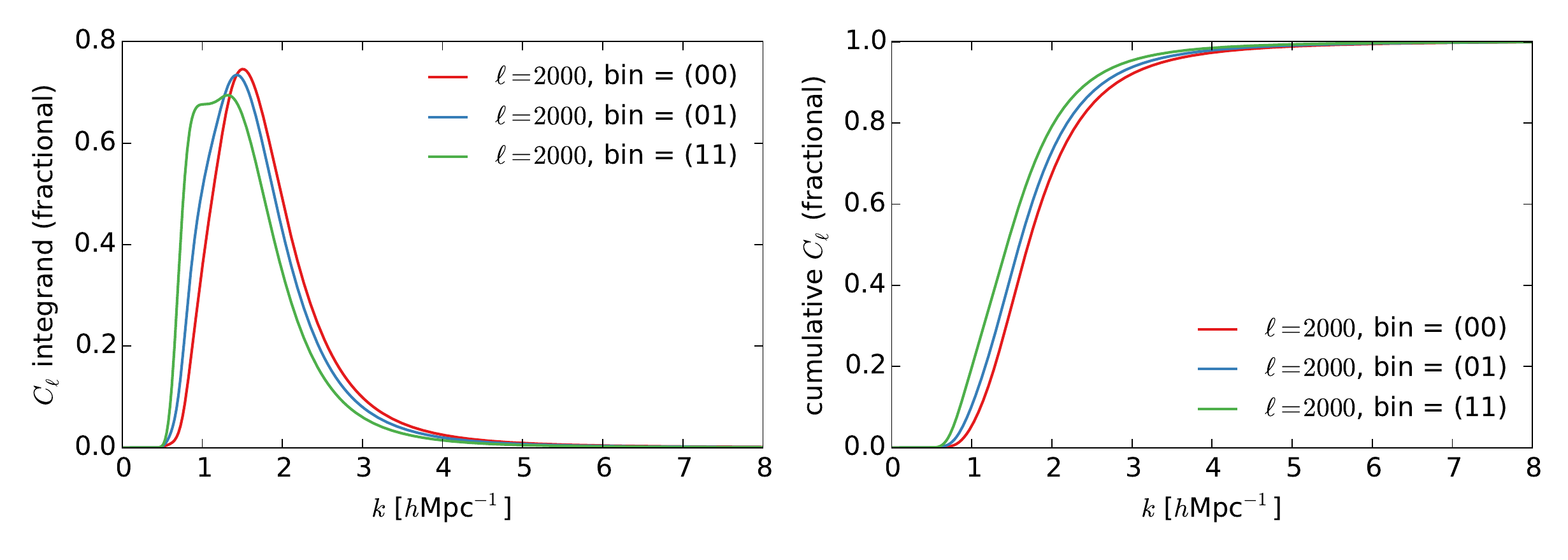}
\caption{Left: Fractional integrand of the tomographic angular power spectrum coefficients $C^\mathrm{EE}_{\ell,\alpha\beta}$ as a function of $k$, showing explicitly the contribution from different scales to the angular power spectra at the maximum $\ell$ probed in this analysis ($\ell = 2000$). Right: The corresponding (fractional) integrals for $C^\mathrm{EE}_{\ell,\alpha\beta}$ as a function of the upper integration limit (where the $C^\mathrm{EE}_{\ell,\alpha\beta}$ is obtained in the limit $k\rightarrow \infty$). Both panels show that by choosing $\ell \leq 2000$ we are effectively removing scales $k < 5h/\mathrm{Mpc}$ from the analysis, whilst $k > 1.5h/\mathrm{Mpc}$ still make significant contribution at higher multipoles. This statement is robust to the range of $\Lambda$CDM parameters allowed under our chosen prior (\cf, \S \ref{sec:cosmo_models})}
\label{fig:kernels}
\end{figure*}
\subsection{Model selection for cosmological Gibbs samplers}
\label{sec:model_selection_gibbs}
When comparing competing models, we would like to compute the odds ratio and perform formal Bayesian model-comparison. Comparing two models $\mathcal{M}_A$ and $\mathcal{M}_B$, the odds ratio is given by
\begin{align}
\label{odds}
\mathcal{O}_{AB} = \frac{P(\data | \mathcal{M}_A)}{P(\data | \mathcal{M}_B)}\f{P(\mathcal{M}_A)}{P( \mathcal{M}_B)}
\end{align}
where $P(\data | \mathcal{M})$ is the Bayesian evidence (or marginal likelihood) for a given model, and the prior odds ratio $P(\mathcal{M}_A)/P( \mathcal{M}_B)$ defines our prior relative belief in model $A$ versus model $B$ -- this is typically set to unity if neither model is preferred \emph{a priori}, in which case the odds ratio reduces to the Bayes factor $\mathcal{K}_{AB} = P(\data | \mathcal{M}_A)/P(\data | \mathcal{M}_B)$. A Bayes factor $\mathcal{K}_{AB} > 1$ indicates preference for model $A$ over model $B$, and vice versa for $\mathcal{K}_{AB} < 1$. In this work we follow the quantitative interpretation scheme of \citet{Kass1995} for interpreting the Bayes factors.

In order to compute the Bayes factor we need to evaluate the evidence for each model, \ie, the integral of the likelihood $P(\data | \bs\theta, \mathcal{M})$ under the prior:
\begin{align}
\label{evidence}
P(\data | \mathcal{M}) = \int P(\data | \bs\theta, \mathcal{M})P(\bs\theta|\mathcal{M})d\bs\theta.
\end{align}
Widely used Monte Carlo integration schemes for computing the evidence (such as nested sampling, \citealp{Skilling2006}) typically involve direct likelihood evaluations, often computing the evidence and generating samples from the likelihood (or posterior) simultaneously. In our case, direct likelihood calculations are prohibitively expensive, requiring $\sim n_\mathrm{pix}\times n_\mathrm{pix}$ matrix inversions. Gibbs sampling is conceptually different from other MCMC schemes and avoids this problem altogether since we never actually evaluate the likelihood (or posterior) directly, rather iteratively drawing samples from a series of conditional densities. However, since it doesn't provide the evidence as a byproduct we need to seek an alternative way of accessing the evidence.

\citet{Chib1995} suggest a simple method for computing the evidence from a set of Gibbs samples. Suppose a model has interesting parameters $\bs\theta$ and some latent variables $\field$ that are marginalised over. From Bayes theorem, the evidence is related to the prior, posterior and likelihood by
\begin{align}
\label{chib_identity}
P(\data | \mathcal{M}) = \f{P(\data | \bs\theta, \mathcal{M})P(\bs\theta | \mathcal{M})}{P(\bs\theta | \data, \mathcal{M})},
\end{align}
hence if we can compute the likelihood, prior and (correctly normalised) posterior at a single point, we are able to straightforwardly estimate the evidence. The prior and likelihood are typically readily available (although likelihood calculations may be expensive in practice), and the posterior density can be estimated from the Gibbs samples $\{\bs\theta, \field\}$; see \citet{Chib1995} for optimal estimation of the posterior density at a point from a set of Gibbs samples using the Blackwell-Rao estimator.

In the case when we are comparing two models that are nested, computing the Bayes factor is greatly simplified. Suppose model $\mathcal{M}_A$ has parameters $\bs\theta$ and model $\mathcal{M}_B$ has parameters $\bs\theta$ plus some additional parameters $\bs\phi$; if $\mathcal{M}_A$ is equivalent to $\mathcal{M}_B$ with the additional parameters fixed to some value $\bs\phi = \bs\phi_0$, models $\mathcal{M}_A$ and $\mathcal{M}_B$ are said to be nested. Assuming the same prior on the common parameters $\bs\theta$ under the two models, the Bayes factor $\mathcal{K}_{AB}$ reduces to \citep{Dickey1971}
\begin{align}
\label{savage_dickey}
\mathcal{K}_{AB} = \frac{P(\bs\phi_0 | \data,\mathcal{M}_B )}{P(\bs\phi_0 | \mathcal{M}_B)}.
\end{align} 
Hence for nested models (under the same prior), provided we can estimate the marginal posterior density of the new parameters at $\bs\phi = \bs\phi_0$ we can compute the Bayes factor by Eq. \eqref{savage_dickey} -- this is referred to as the Savage-Dickey Density Ratio. The marginal posterior density $P(\bs\phi_0 | \data,\mathcal{M}_B )$ can be estimated quickly and accurately from a set of MCMC samples using a histogram, kernel density estimate \citep{Silverman1986}, or some other density estimation scheme (provided the dimensionality of $\bs\phi$ is not prohibitively large).

For nested models with a different prior on the common parameters under the two models, we can compute the Bayes factor from Eq. \eqref{chib_identity},
\begin{align}
\label{chib_ratio}
\mathcal{K}_{AB} =\frac{P(\bs\theta, \bs\phi_0 | \data,\mathcal{M}_B )P(\bs\theta | \mathcal{M}_A) }{P(\bs\theta, \bs\phi_0 | \mathcal{M}_B)P(\bs\theta | \data,\mathcal{M}_A )},
\end{align}
where again the posterior densities can be estimated from the MCMC samples, and under the assumption of nested models the likelihood terms from Eq. \eqref{chib_identity} cancel exactly, circumventing the need for brute force likelihood evaluation. In this case we must estimate the posterior densities over the full parameter spaces $\bs\theta$ and $(\bs\theta, \bs\phi)$.

In this work we only consider models that are (approximately) nested with $\Lambda$CDM and use either the Savage Dickey Density Ratio or Eq. \ref{chib_ratio} for fast and accurate evaluation of Bayes factors, using Gaussian kernel density estimates to estimate the required posterior densities.

\section{Cosmological models and implementation}
\label{sec:implementation}
\subsection{Physical scales}
The $E$-mode tomographic power spectra $C^\mathrm{EE}_{\ell, \alpha\beta}$ are functionals of the matter power spectrum, where each $\ell$ mode for a given redshift bin combination probes a range of $k$-scales of the matter density field (with high $\ell$s and low redshifts probing higher $k$ modes). In this analysis, we restrict the $\ell$ range to limit contributions from theoretically uncertain baryonic effects in the matter power spectrum at high-$k$. To this end, we consider $\ell \leq 2000$ for both tomographic bins; this ensures that no power spectrum coefficient $C^\mathrm{EE}_{\ell, \alpha\beta}$ picks up more than $\sim 5\%$ contribution from $k > 5h/\mathrm{Mpc}$ (for any allowed cosmology under the prior), so any anticipated corrections due to baryonic physics are expected to have a sub-percent level effect on the highest $C^\mathrm{EE}_{\ell, \alpha\beta}$ coefficients included in our analysis.

Fig. \ref{fig:kernels} shows the contribution to the $C^\mathrm{EE}_{\ell, \alpha\beta}$ coefficients as a function of $k$ at $\ell = 2000$ (for a fiducial \emph{Planck} 2015 cosmology). We can calculate the contribution from different $k$-scales to the $C^\mathrm{EE}_{\ell,\alpha\beta}$ as follows: The tomographic $E$-mode power spectra can be written as integrals (over $k$) of the matter power spectrum with a geometric kernel (\cf, Eq. \eqref{limber_power}):
\begin{align}
C^\mathrm{EE}_{\ell,\alpha\beta} &= \ell^{-1}\int dk\;w_\alpha(\ell/k)w_\beta(\ell/k)P_\delta(k; \ell/k) \nn \\
&\equiv \int_0^\infty I^\mathrm{EE}_{\ell,\alpha\beta}(k)dk.
\end{align}
To explore the contribution to the angular power spectra as a function of $k$, we can look at the integrand $I^\mathrm{EE}_{\ell,\alpha\beta}(k)$. The integrand is plotted in Fig. \ref{fig:kernels} for the $\ell = 2000$ mode for all three cross power spectra (left panel) with the corresponding cumulative integral $\int_0^k dk' I^\mathrm{EE}_{\ell,\alpha\beta}(k')$ (right panel). This shows clearly that the total contribution to the $C^\mathrm{EE}_{\ell,\alpha\beta}$ for the $\ell = 2000$ mode picks up almost no contributions from $k > 5h/\mathrm{Mpc}$ where the theoretical uncertainties are largest --- this statement is robust to the range of $\Lambda$CDM parameters allowed under our prior (\cf, \S \ref{sec:cosmo_models}). For lower $\ell$-modes, the integrands have similar shapes but are translated towards lower $k$.

Compared to previous CFHTLenS analyses, our scale cuts are less conservative than the $k < 1.5h/\mathrm{Mpc}$ cut employed in the 3D weak lensing analyses of \citet{Kitching20143D}, but slightly more conservative than their $k < 5h/\mathrm{Mpc}$ cut since the 3D power spectra have narrower kernels in $k$ compared to tomography. Our scale cuts are generally more conservative than correlation function analyses that suffer from greater mixing of angular scales \citep{Joudaki2016, Heymans2013, Benjamin2013, Kilbinger2013}.
\subsection{Band powers}
For low signal-to-noise small survey-area analyses, the map-power spectrum Gibbs sampler benefits from binning $\ell$ modes together into band powers \citep{Larson2007, Eriksen2006}; by binning $\ell$ modes together into a set of band-powers, we are increasing the effective signal-to-noise of a reduced set of power spectrum coefficients, hence improving the sampling efficiency. However, this binning requires the assumption of a fiducial band-power shape -- this is ultimately an approximation and should be avoided in an optimal map-power spectrum analysis aimed at high quality cosmological parameter inferences. In this work we reserve cosmological parameter inference for the map-cosmology inference scheme --- as such, in our map-power spectrum inference implementation we allow ourselves to bin $\ell$ modes into a number of broad band-powers of width $\Delta\ell = 200$ (with the exception of the lowest bin which spans $\ell = 30 - 200$).

We specify a fiducial shape inside each $E$-mode band-power $C^\mathrm{F}_\ell = C^\mathrm{EE}_{\ell,00}$ corresponding to low redshift bin auto-power computed for a \emph{Planck} 2015 \citep{Planck2015XIII} cosmology, \ie, the power spectrum inside band $\mathcal{B}$ takes the form
\begin{align}
\cov_{\ell} = \cov_\mathcal{B} C^\mathrm{F}_\ell.
\end{align}
We assume flat band powers for the $B$-modes. Previous CFHTLenS analyses have found no evidence for parity violating $EB$-correlations, so we will neglect $EB$ correlations in this study but note that recovery of $EB$ power using the Bayesian methods developed here is straightforward \citep{Alsing2016}.

Importantly, note that for the map-cosmology inference approach we do not bin into band powers and treats each $\ell$-mode distinctly --- we will use this `band-power approximation free' approach for the final cosmological parameter inferences presented in this work.
\subsection{Cosmological models}
\label{sec:cosmo_models}
Our baseline model is a flat $\Lambda$CDM cosmology with five free parameters: $\ln (10^{10}A_\mathrm{S})$, $\Omega_\mathrm{m}$, $\Omega_\mathrm{b}$, $h$ and $n_s$, denoting the amplitude of scalar fluctuations, matter and baryon density parameters, Hubble constant and scalar spectral index respectively. Following \citet{Planck2015XIII} we include two massless and one massive neutrino with $m_\nu = 0.06\mathrm{eV}$ in our baseline model (\ie, a normal mass hierarchy and single dominant mass eigenstate).

In the first extension to $\Lambda$CDM we consider a model with three degenerate massive neutrinos, with the total neutrino mass $\sum m_\nu$ as an additional (sixth) free parameter (neglecting the small differences in mass expected from the observed mass splittings). We will denote this model $\Lambda$CDM$+m_\nu$. The detection of neutrino oscillations has firmly established that neutrinos have mass, constraining the differences of the square masses between neutrino species $\Delta m^2_{13}$ and $\Delta m^2_{23}$ and establishing a minimum total neutrino mass $\sum m_\nu \gsim 0.06 \mathrm{eV}$ (see \eg, \citet{Forero2012} for a review). Whilst our baseline model assumes a normal mass hierarchy with minimal mass (and a single dominant mass eigenstate), constraints on $\Delta m^2_{13}$ and $\Delta m^2_{23}$ alone allow for many other scenarios, including a degenerate hierarchy with $\sum m_\nu \gsim 0.1 \mathrm{eV}$ or an inverted hierarchy (see \eg, \citealp{Lesgourgues2006}). At the present time there are no compelling theoretical or empirical reasons to prefer any of these possibilities over another, and as such allowing for non-minimal neutrino masses is one of the most well-motivated extensions to the baseline model. Constraints on total neutrino mass from cosmology are already significantly stronger than those from tritium beta decay experiments, with CMB observations combined with baryon acoustic oscillations constraining $\sum m_\nu < 0.17 \mathrm{eV}$ (95\%) \citep{Planck2015XIII} compared to $\sum m_\nu < 6 \mathrm{eV}$ (95\%) from beta decay experiments \citep{Drexlin2013} (although some caution is deserved, since the cosmology-derived limits are highly model dependent). Upper and lower limits on the total neutrino mass from beta decay and neutrino oscillation experiments nonetheless provide a well-motivated prior range for the total neutrino mass $\sum m_\nu \in \left[0.06, 6\right]\;\mathrm{eV}$, adopted in this work.

In the second extension, we include the possibility of a redshift-dependent bias on the CFHTLenS photo-$z$ measurements. In an analysis of cross-correlations between photometric CFHTLenS galaxies and spectroscopic BOSS galaxies in overlapping regions, \citet{Choi2015} find evidence for a redshift dependent bias in the CFHTLenS photo-$z$s. In a 3D cosmic shear power spectrum analysis, \citet{Kitching2016} also find suggestions of a significant photo-$z$ bias (assuming a fixed \emph{Planck} 2015 cosmology). \citet{Joudaki2016} find some support for photo-$z$ biases too, where including a model for photo-$z$ bias improved their reported tension with \emph{Planck} and flexible photo-$z$ biases (combined with other systematics) were preferred by the CFHTLenS data. In light of these studies, models including photo-$z$ biases are an important and well-motivated extension to the baseline model for CFHTLenS. Following \citet{Kitching2016}, we model the photo-$z$ as having a linear redshift dependent bias $\Delta z (z_\mathrm{phot}) = p_2(z_\mathrm{phot}-p_1)$, so the tomographic redshfit distributions shift according to $n_\alpha(z)\rightarrow n^\alpha(z - \Delta z(z))/\int n^\alpha(z - \Delta z(z))dz$ (taking care to renormalise the distributions after the shift). We treat $p_2$ and $p_1$ as additional free parameters with flat priors over $p_1\in \left[-0.5, 0.5\right]$ and $p_2\in \left[-0.5, 0.5\right]$ and denote this model  $\Lambda$CDM$+\Delta z$.

We do not attempt to model baryonic suppression of the power spectrum on small scales. This may be justified since we cut small scales quite aggressively from our analysis, only considering $\ell \leq 2000$. Furthermore \citet{Kitching2016} find no strong evidence for baryonic suppression in the CFHTLenS data when analysed with a \emph{Planck} 2015 prior on the cosmology, and \citet{Kohlinger2016} (employing the same tomographic binning and a similar scale-cut to our work) also find that models with baryonic suppression are not preferred by the CFHTLenS data, finding an amplitude of baryon suppression consistent with zero. We also neglect intrinsic alignments in our analysis, since for the broad redshift bins used in this analysis the intrinsic alignment contamination to the power spectra is expected to be at the percent-level or smaller \citep{Benjamin2013, Sifon2015, Kohlinger2016}.

Throughout this work we use \textsc{camb}\footnote{http://camb.info version 2015} to compute the non-linear matter power spectrum and assume priors as outlined in the following section.

\subsection{Priors}
\label{sec:priors}
\subsubsection*{$\Lambda$CDM and neutrino mass priors}
Under all of the models considered in this work we assume flat priors for the $\Lambda$CDM cosmological parameters over the following ranges: $\ln(10^{10}A_\mathrm{S}) \in \left[0, 20\right]$, $\Omega_\mathrm{m}\in \left[0, 1\right]$, $\Omega_\mathrm{b}\in \left[0, 0.1\right]$, $h\in \left[0.4, 1.0\right]$ and $n_s\in \left[0.7,1.3\right]$. For the $\Lambda$CDM$+m_\nu$ model we take a flat prior on the total neutrino mass $\sum m_\nu \in \left[0.06, 6\right]\mathrm{eV}$, where the upper limit is motivated by tritium beta-decay experiments \citep{Drexlin2013} and the lower limit from neutrino oscillations.
\subsubsection*{Photo-$z$ bias priors}
For the photo-$z$ bias model $\Lambda$CDM$+\Delta z$ we consider two priors on the photo-$z$ bias parameters $p_1$ and $p_2$. Firstly, we consider a broad flat prior $p_1\in \left[-0.5, 0.5\right]$ and $p_2\in \left[-0.5, 0.5\right]$, conservatively assuming little \emph{a priori} knowledge of the bias parameters. Secondly, we use the CFHTLenS photo-$z$ bias constraints from \citet{Choi2015} to construct a well-motivated informative prior over $p_1$ and $p_2$ based on independent measurements.

\citet{Choi2015} performed a cross-correlation analysis of CFHTLenS photometric galaxies with BOSS spectroscopic sources to constrain a mean photo-$z$ bias $\Delta z$ in five redshift bins -- their measurements are summarised in Table \ref{tab:choi}. We use their measurements to constrain our linear photo-$z$ bias parameters $p_1$ and $p_2$ as follows: The mean photo-$z$ bias in redshift bin $\alpha$ for a given value of $p_1$ and $p_2$ is given by $\bar{\Delta} z_\alpha(p_1, p_2) = \sum_{z_B\in \alpha}p_2(z_B - p_1)$, where the sum is performed over the \textsc{bpz} photo-$z$s for the CFHTLenS sources in each redshift bin. Assuming independent Gaussian errors on the $\Delta z$ measurements we can construct the posterior distribution of $p_1$ and $p_2$ given their measurements (assuming wide uniform priors):
\begin{align}
P(p_1, p_2 | \{\Delta z_\alpha\})\propto \prod_\alpha \exp\left[-\f{1}{2}(\bar{\Delta} z_\alpha(p_1, p_2) - \Delta z_\alpha)^2/\sigma_{\alpha}^2\right].
\end{align}
The posterior on $p_1$ and $p_2$ from the \citet{Choi2015} measurements is well approximated by a bivariate Gaussian with mean and covariance given in Table \ref{tab:choi}. These constraints provide a well-motivated informative Gaussian prior on the photo-$z$ bias parameters from an independent analysis.
\begin{table}
\small
\begin{center}
\caption{Measurements of the mean photo-$z$ bias in five redshift bins from \citet{Choi2015} and the resulting constraints on photo-$z$ bias parameters $p_1$ and $p_2$ for the model $\Delta z = p_2(z - p_1)$, assuming independent Gaussian errors on the measurements and broad uniform priors for $p_1$ and $p_2$. }
\begin{tabular}{cc}
\hline
Measurements from \citet{Choi2015} & \\
\hline\\
$\Delta z_1\;\;0.15 < z_B < 0.29$ & $-0.037^{\scriptscriptstyle +0.009}_{\scriptscriptstyle -0.010}$ \tblspacer \\
$\Delta z_2\;\;0.29 < z_B < 0.43$ & $-0.016^{\scriptscriptstyle +0.008}_{\scriptscriptstyle -0.008}$ \tblspacer \\
$\Delta z_3\;\;0.43 < z_B < 0.57$ & $0.007^{\scriptscriptstyle +0.006}_{\scriptscriptstyle -0.006}$\tblspacer  \\
$\Delta z_4\;\;0.57 < z_B < 0.70$ & $0.049^{\scriptscriptstyle +0.010}_{\scriptscriptstyle -0.010}$ \tblspacer \\
$\Delta z_5\;\;0.70 < z_B < 0.90$ & $0.036^{\scriptscriptstyle +0.016}_{\scriptscriptstyle -0.013}$ \tblspacer \\
\hline
Resulting constraints on $p_1$ and $p_2$ & \\
\hline \\
mean $p_1$ & 0.45 \\
mean $p_2$ & -0.16 \\
$\mathrm{cov}(p_1, p_1)$ & 0.00064 \\
$\mathrm{cov}(p_1, p_2)$ & -0.0001 \\
$\mathrm{cov}(p_2, p_2)$ & 0.00065 \\
\hline
\end{tabular}
\label{tab:choi}
\end{center}
\end{table}
\subsection{Comparison with Planck}
\label{sec:planck_definition}
When comparing results from CFHTLenS against cosmic microwave background measurements from \emph{Planck} \citep{Planck2015XIII}, we consider combined CMB temperature and low-$\ell$ polarization constraints, which we will refer to simply as \emph{Planck} 2015. We do not consider small scale polarization or CMB lensing measurements in this analysis.
\begin{figure*}
\centering
\includegraphics[width = 17.5cm]{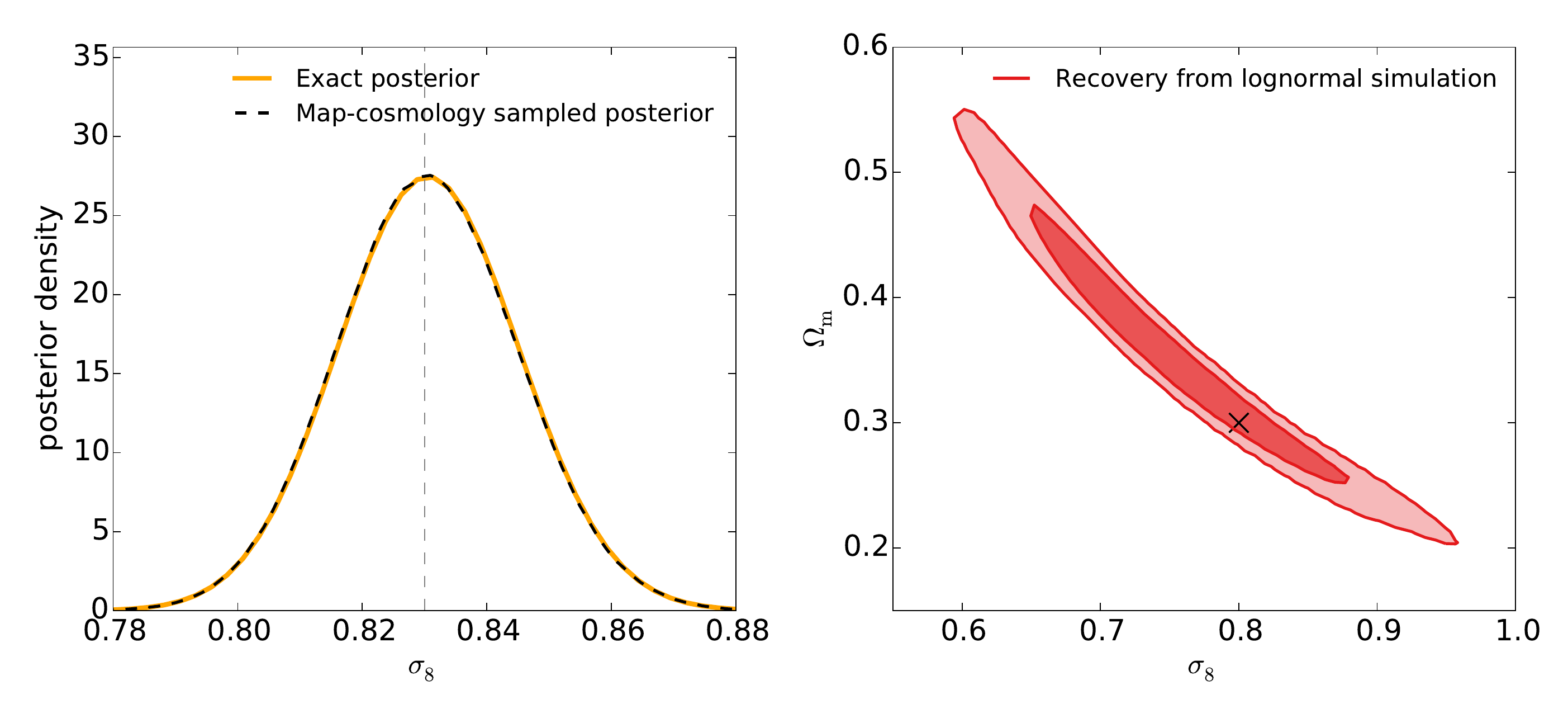}
\caption{Demonstration of the map-cosmology inference scheme on simulated data. Left: Recovered posterior distribution of $\sigma_8$ using the map-cosmology sampling scheme (black dashed) compared to the corresponding exact posterior density (orange), from a Gaussian shear simulation with no mask and isotropic noise (where comparison with the exact posterior is numerically tractable). The exact posterior is well recovered by the map-cosmology sampler. Right: Recovered posterior distribution of $\sigma_8$ and $\Omega_\mathrm{m}$ from a shear simulation with lognormal lensing statistics and comparable signal-to-noise, mask and survey area to CFHTLenS. The contours represent 68 and 95\% credible intervals and the black cross indicates the input value. The cosmological parameters are well recovered from the simulated data.}
\label{fig:sim_recovery}
\end{figure*}
\section{Results}
\label{sec:results}
\subsection{Demonstration on simulations}
\label{sec:sim_results}
Before analysing the CFHTLenS data, we want to demonstrate the map-cosmology sampling scheme on simulations (the map-power spectrum inference scheme was demonstrated on realistic shear simulations in \citealp{Alsing2016}). To this end, we run the map-cosmology sampling algorithm on two types of simulation: (1) a Gaussian shear simulation with no mask and isotropic noise, for which comparison with an exact (analytically and numerically tractable) posterior is possible, and (2) a (more realistic) lognormal shear simulation with comparable mask, survey-area and signal-to-noise to CFHTLenS.
\subsubsection*{Comparison to exact (analytical) posterior inference}
In the idealised case of isotropic noise $\noise = \sigma^2\mathbf{I}$, trivial survey geometry and no mask (and assuming Gaussian fields), the posterior distribution of the cosmological parameters can be readily computed exactly from the data:
\begin{align}
\label{exact_posterior}
P(\bs\theta | \data) \propto \prod \f{1}{|\cov_\ell(\bs\theta) + \noise_\ell |^\f{n_\ell}{2}} e^{-\f{1}{2}\mathrm{tr}\left[\mathbf{D}_\ell(\cov_\ell(\bs\theta) + \noise_\ell)^{-1}\right]} \times P(\bs\theta),
\end{align}
where $n_\ell = (\sum_{|\bs\ell| = \ell})$ is the number of Fourier modes contributing to mode $\ell$, $\mathbf{D}_\ell = \sum_{|\bs\ell| = \ell} \data_{\bs\ell}^{\phantom\dagger}\data_{\bs\ell}^\dagger$ (i.e., the outer product of Fourier modes of the data for a given $\ell$ mode) and $\noise_\ell = \sigma^2 \mathbf{I}_{n_{bins}\times n_{bins}}$ is the noise covariance for mode $\ell$ (identical for all modes under the assumption of isotropic noise). This idealised case allows us to compare the Bayesian sampling schemes against an analytically and numerically accessible exact posterior density.

To demonstrate the map-cosmology sampling scheme in this idealised case, we generated a Gaussian shear field realisation over a $150$ square degree patch, taking the tomographic redshift distributions constructed from the CFHTLenS data (\cf, Fig. \ref{fig:cfhtlens_photo_z_dist}), assuming a \emph{Planck} 2015 $\Lambda$CDM cosmology and pixelised into $87\times87$ pixels to restrict scales to $\ell \leq 2000$. We add isotropic Gaussian noise assuming an ellipticity dispersion of $\sigma_\epsilon = 0.279$ per component and the same mean source density as CFHTLenS (averaged into a constant noise-covariance value across both tomographic slices). The map-cosmology sampler was then run on the simulated noisy shear maps to recover $\sigma_8$\footnote{In the idealised case of isotropic noise and no mask, the messenger field (\cf, \S \ref{sec:bayes_schemes}) is superfluous. However, to test all elements of the algorithm, we keep the messenger field sampling step and partition the noise covariance $\noise = \sigma^2\mathbf{I}$ according to $\isotropic = \tau\mathbf{I}$ and $\aniso = \noise - \isotropic$ with $\tau = 0.9\sigma^2$.}. The recovered posterior density for $\sigma_8$ and the exact posterior computed from Eq. \eqref{exact_posterior} are shown in Fig. \ref{fig:sim_recovery} (left panel); the exact posterior is clearly well recovered by the map-cosmology sampling scheme. 

As far as the Bayesian sampling scheme is concerned, extension to anisotropic noise and non-trivial mask is conceptually and numerically simple; the isotropic and anisotropic noise covariances $\isotropic$ and $\aniso$ are just replaced accordingly (\cf, \S \ref{sec:bayes_schemes}). Both of these matrices will still be simple diagonal matrices in the pixel domain\footnote{The noise covariances are diagonal assuming uncorrelated pixel noise. Note that intrinsic alignments should be treated as an additional contribution to the signal covariance $\cov$ rather than as correlations in the pixel-noise covariances, \cf, \citet{Alsing2016}.}, so anisotropic noise and mask are not expected to introduce any further numerical complications. Hence, recovery of the exact posterior density in the simplified case shown in Fig. \ref{fig:sim_recovery} (left) is a strong test of the algorithm.
\subsubsection*{Demonstration on lognormal shear simulations}
In addition to the idealised case of Gaussian fields with isotropic noise and no mask, we would like to demonstrate the map-cosmology sampler on a more realistic shear simulation with comparable characteristics to the CFHTLenS data. To this end, we construct a lognormal shear field simulation, with anisotropic noise and a complicated mask (comparable to the CFHTLenS data).

We generate a lognormal random shear field realisation from a set of input power spectra as follows: Firstly, the input convergence power spectrum is Fourier (Hankel) transformed to real-space convergence correlation functions $C^\mathrm{EE}_{\ell,\alpha\beta} \rightarrow \xi_{\alpha\beta}(\theta)$. These correlation functions are then transformed according to $\bs\xi \rightarrow \ln\left[1 + \bs\xi/a^2\right]$; this is the closed-form relation between the correlation functions of a lognormal field and that of the associated logarithmically transformed Gaussian random field (see e.g., \citealp{Hilbert2011, Joachimi2011a}). The lognormal parameter $a$ is fixed to $a = 0.012$, consistent with findings from the Millennium simulations \citep{Hilbert2011}. The correlation functions of the Gaussianised (i.e., log-transformed) fields are then transformed back to power spectra and from this set of Gaussianised power spectra, we generate correlated Gaussian random fields for the two tomographic slices. The resulting fields are then exponentiated to produce lognormally distributed convergence fields whose two-point statistics (and correlations) match the input power spectra. Finally, shear fields are derived from the convergence fields from their simple relation in Fourier space: $\gamma_{\bs\ell} = e^{2i\varphi_{\bs\ell}}\kappa_{\bs\ell}$, with $\tan\varphi_{\bs\ell} = \ell_x/\ell_y$. 

As before, we simulate a $150$ square degree patch pixelized into $87\times 87$ pixels to restrict the analysis to modes with $\ell \leq 2000$ and assume a fiducial \emph{Planck} 2015 cosmology. We add (anisotropic) Gaussian noise to each pixel, assuming a mean source density of $15$ galaxies per square arc-minute in each redshift bin, with Poisson distributed source numbers per pixel and an ellipticity dispersion of $\sigma_e = 0.279$ per component. To simulate the effects of the survey mask, we mask out ten randomly positioned one square degree patches to mimic exclusion of ``bad fields", and $20$ circular patches of radius $6$ arc-minutes to mimic masking of (larger) point sources.  

We applied the map-cosmology inference scheme to the lognormal simulations and recover cosmological parameters. We ran ten independent MCMC chains of $\num[group-separator={,}]{100000}$ steps, ensuring a Gelman-Rubin statistic of $R < 1.03$ for all $\Lambda$CDM parameters \citep{Gelman1992}. The marginal posterior density in $\sigma_8$ and $\Omega_\mathrm{m}$ is shown in Fig. \ref{fig:sim_recovery} (right panel) from a single lognormal simulation realisation; the cosmological parameters are well recovered from the simulated data. We leave an extensive code verification and stress-testing program using a very large number of simulations to future work, but nonetheless Fig. \ref{fig:sim_recovery} demonstrates the map-cosmology inference scheme is able to recover cosmological parameters from realistic weak lensing simulations.
\begin{figure*}
\centering
\includegraphics[width = 18cm]{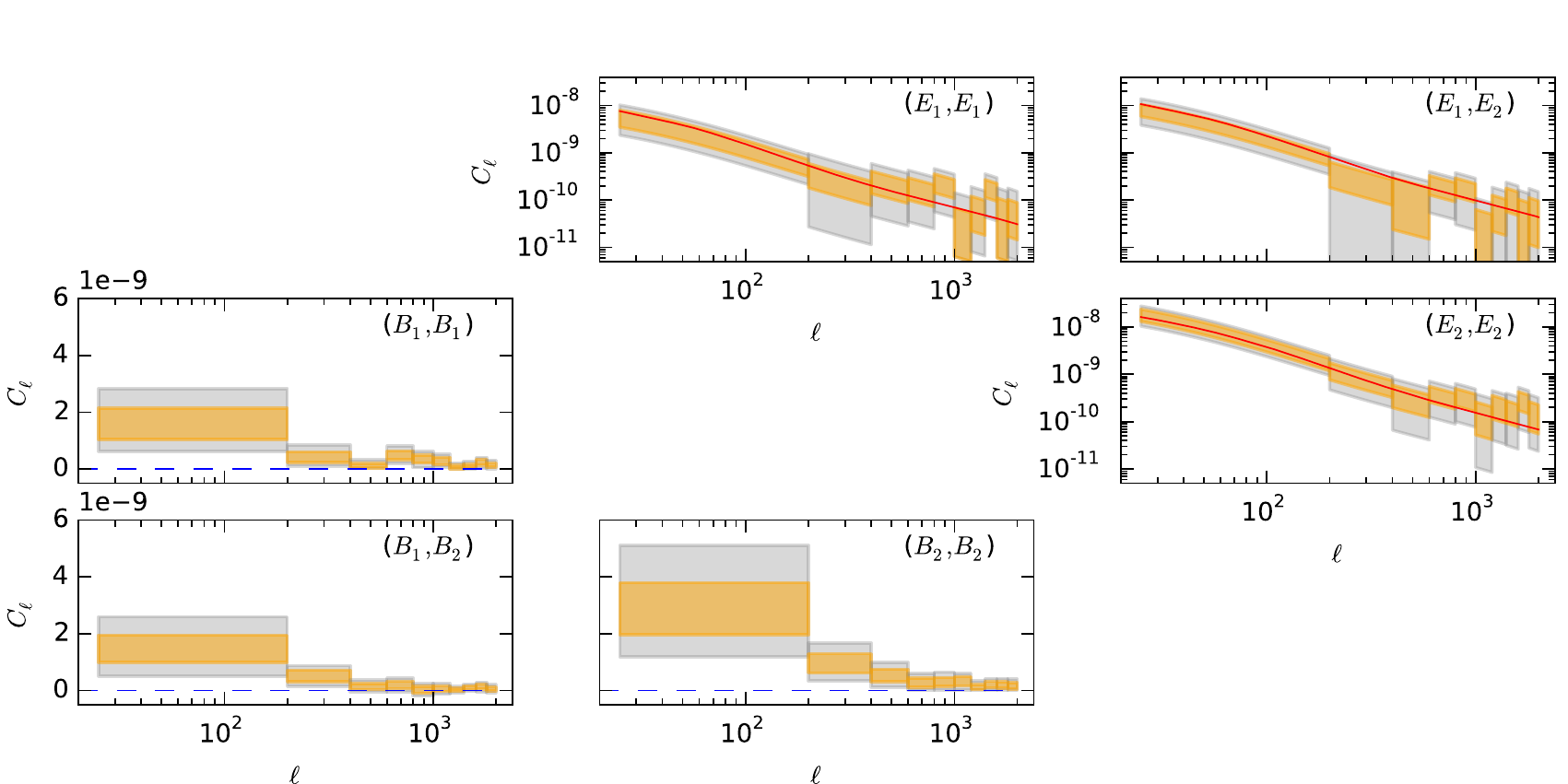}
\caption{Recovered posteriors for the $E$- and $B$-mode tomographic power spectra from CFHTLenS, summarised by $68\%$ (orange) and $95\%$ (grey) credible intervals. The best-fit (maximum posterior) $\Lambda$CDM model is shown in red (obtained from the map-cosmology sampling scheme applied to the CFHTLenS data, c.f., \S \ref{sec:cfhtlens_cosmology}).}
\label{fig:cfhtlens_ebmode}
\end{figure*}
\subsection{CFHTLenS power spectra}
We ran the map-power spectrum inference algorithm on the pixelized CFHTLenS maps described in \S \ref{sec:maps}, running ten independent Gibbs chains of $\num[group-separator={,}]{500000}$ samples each. Convergence was assessed by ensuring the Gelman-Rubin statistic $R < 1.03$ in all parameters.

The recovered $E$-mode power spectra are shown in Fig. \ref{fig:cfhtlens_ebmode}, where the posterior inference is summarised by 68 (orange) and 95\% (grey) credible regions and the best-fit $\Lambda$CDM model is shown in red (see \S \ref{sec:cfhtlens_cosmology}); the inferred $E$-mode power spectra are well described by the $\Lambda$CDM fit. These band-power posteriors can in principle be used to infer cosmological parameters, provided we can $(a)$ accurately smooth the samples to reconstruct the posterior density $P(\cov | \data)$, and $(b)$ make an appropriate mapping from theory predicted $\{C_\ell\}$ to band-power coefficients. We leave cosmological parameter inference from the power spectrum samples to future work, appealing to the map-cosmology inference scheme (that circumvents both of these steps) for inferring cosmological parameters in this paper.

The recovered $B$-mode posteriors are summarized in Fig. \ref{fig:cfhtlens_ebmode}. On scales $\ell \gsim 400$ the $B$-modes are broadly consistent with zero, with some scatter. Also note the requirement that the $B$-mode auto-power spectra are positive semi-definite leads to positively skewed posterior densities. On large scales $\ell < 400$ there appears to be a non-negligible $B$-mode signal. This is in line with \citet{Asgari2016} who report a (statistically) significant detection of $B$-modes in CFHTLenS on scales $\theta > 40'$, which roughly corresponds to $\ell \lsim 300$. 

Note that the $B$-mode inference summarised in Fig. \ref{fig:cfhtlens_ebmode} assumed a uniform prior $P(\cov)\propto \mathrm{const}.$ over the $B$-mode power spectra. This is not necessarily the most agnostic (uninformative) prior choice, and we would ideally rather take the reference prior $P(\cov)\propto |\cov + \mathbf{N}|^{-(p+1)/2}$ \citep{Daniels1999}. However, owing to the fact that $\cov$ and $\mathbf{N}$ are not sparse in the same basis\footnote{Recall that $\cov$ is sparse in Fourier space, whilst $\mathbf{N}$ is sparse in pixel space.} this prior is prohibitively expensive to compute, involving an $\sim n_\mathrm{pix}\times n_\mathrm{pix}$ matrix determinant at each sampling step. However, in cases where the $B$-mode signal is constrained to be well below the noise-level $\mathbf{N}\gg\cov^\mathrm{BB}$ the effect of assuming a uniform prior over the formally uninformative reference prior may be small. A thorough analysis of the $B$-mode posteriors should perform model-selection on $E$- and $B$-mode models versus $E$-mode only, with an appropriately motivated or uninformative prior on the $B$-modes. Alternatively, one could fit the recovered $B$-mode power spectra with a parametrised model (alongside the cosmological parameters), if a well-motivated model was available. We leave detailed analysis of the cosmic shear $B$-modes in a Bayesian context to future work.

\begin{figure}
\centering
\includegraphics[width = 9cm]{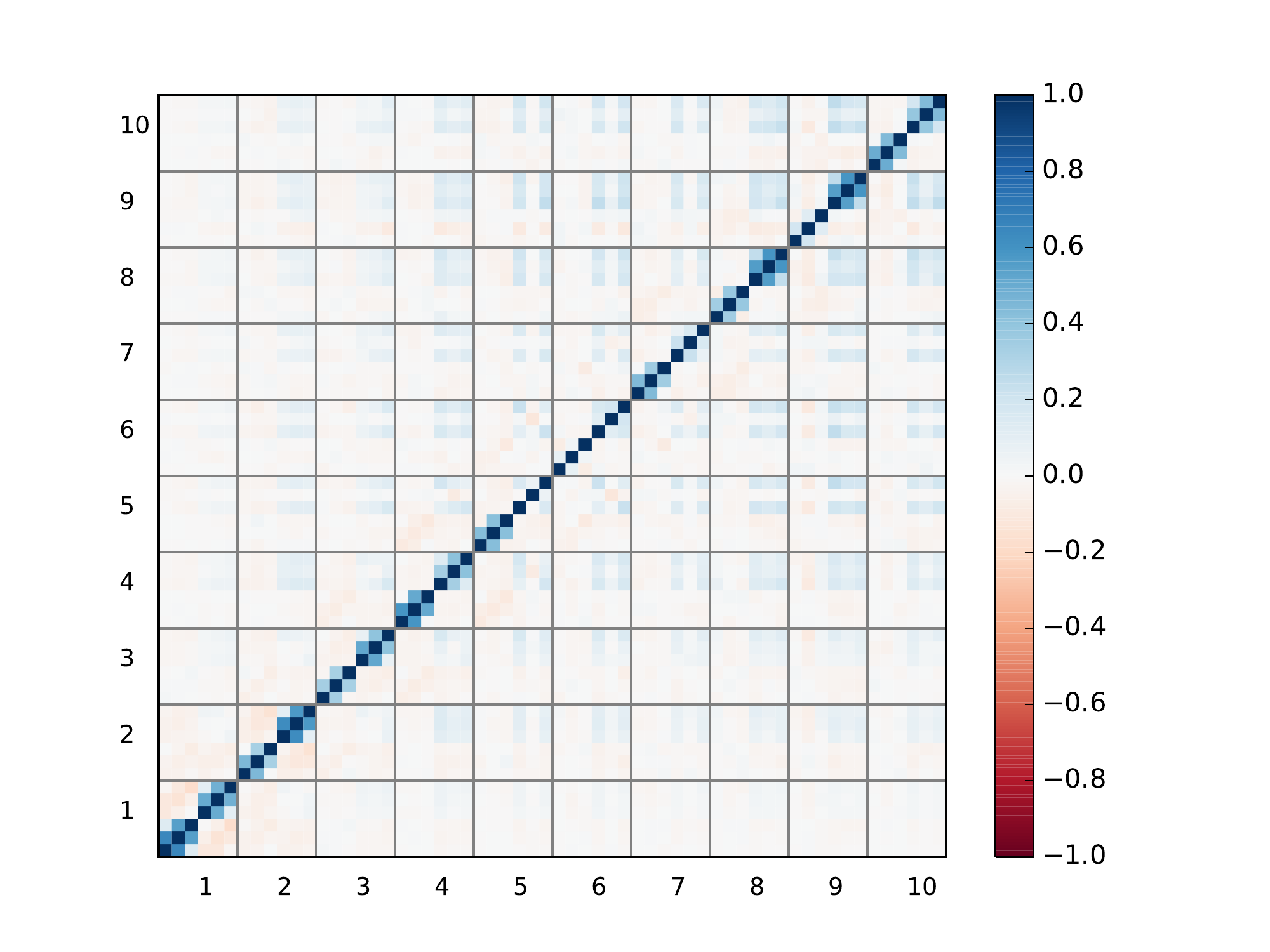}
\caption{Correlation matrix of the posterior band power samples from CFHTLenS. $E/B$-mode band powers are organized into a vector: $\cov = (C^\mathrm{EE}_{\mathcal{B},11}, C^\mathrm{EE}_{\mathcal{B},12}, C^\mathrm{EE}_{\mathcal{B},22}, C^\mathrm{BB}_{\mathcal{B},11}, C^\mathrm{BB}_{\mathcal{B},12}, C^\mathrm{BB}_{\mathcal{B},22}, \dots)$. The correlations between adjacent $E$-mode band powers are typically $\lsim0.1$ and the correlations between the $E$- and $B$-mode inferences are small.}
\label{fig:correlationEB}
\end{figure}

The correlation matrix of the posterior samples organised into a vector $\cov = (C^\mathrm{EE}_{\mathcal{B},11}, C^\mathrm{EE}_{\mathcal{B},12}, C^\mathrm{EE}_{\mathcal{B},22}, C^\mathrm{BB}_{\mathcal{B},11}, C^\mathrm{BB}_{\mathcal{B},12}, C^\mathrm{BB}_{\mathcal{B},22}, \dots)$ is shown in Fig. \ref{fig:correlationEB}, where the grid indicates the ten band-powers. The $3\times 3$ highly correlated blocks along the diagonal represent strong correlations between the three $E$-mode tomographic cross power spectra within each band, as we would expect. There is little ($\lsim 0.1$) correlation between the band powers, so a cosmological analysis of the band-powers (not attempted here) could take the band-powers as being independent to a reasonably good approximation. The correlation between $E$- and $B$-modes is also very small -- this indicates that whilst the presence of $B$-modes on large scales might be alarming (indicating residual unaccounted for systematics in the CFHTLenS data), formally marginalizing over $B$-modes should have a negligible effect on the final parameter inference. Therefore, we are justified (to a good approximation) in ignoring $B$-modes in the map-cosmology inference scheme implementation in this work.
\subsection{CFHTLenS shear maps}
\begin{figure*}
\centering
\includegraphics[width = 17.5cm]{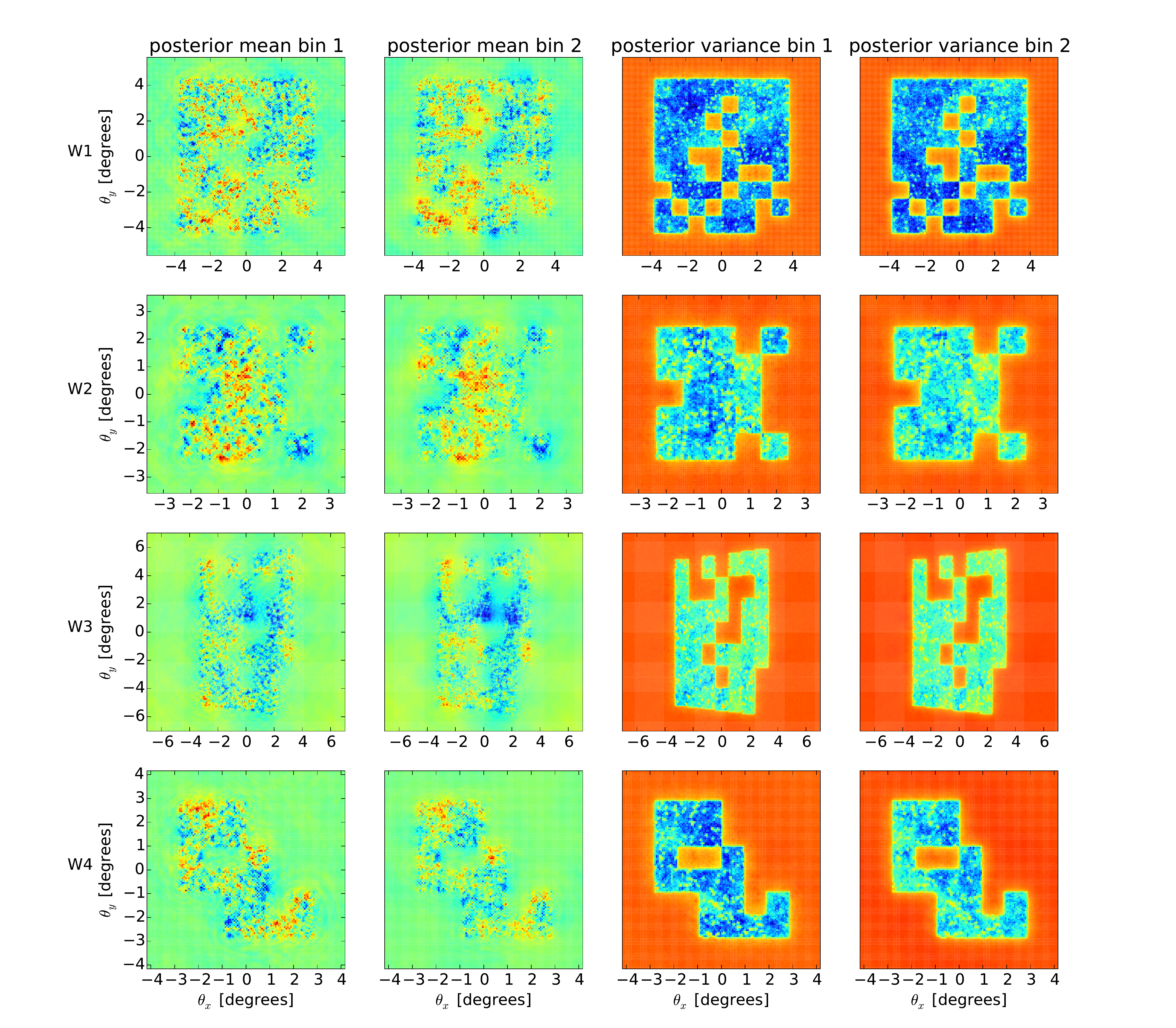}
\includegraphics[width = 17.5cm]{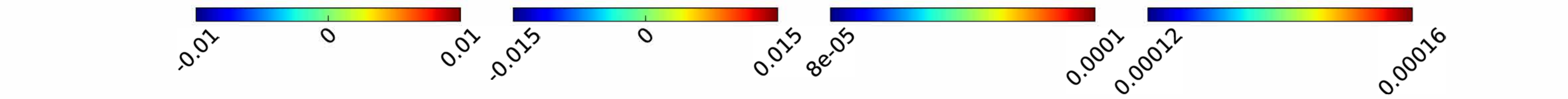}
\caption{Recovered posterior mean and variance for the $\gamma_1$ maps in the four CFHTLenS fields and two tomographic bins. The corresponding $\gamma_2$ maps are shown in Fig. \ref{fig:cfhtlens_map2}}
\label{fig:cfhtlens_map1}
\end{figure*}
\begin{figure*}
\centering
\includegraphics[width = 17.5cm]{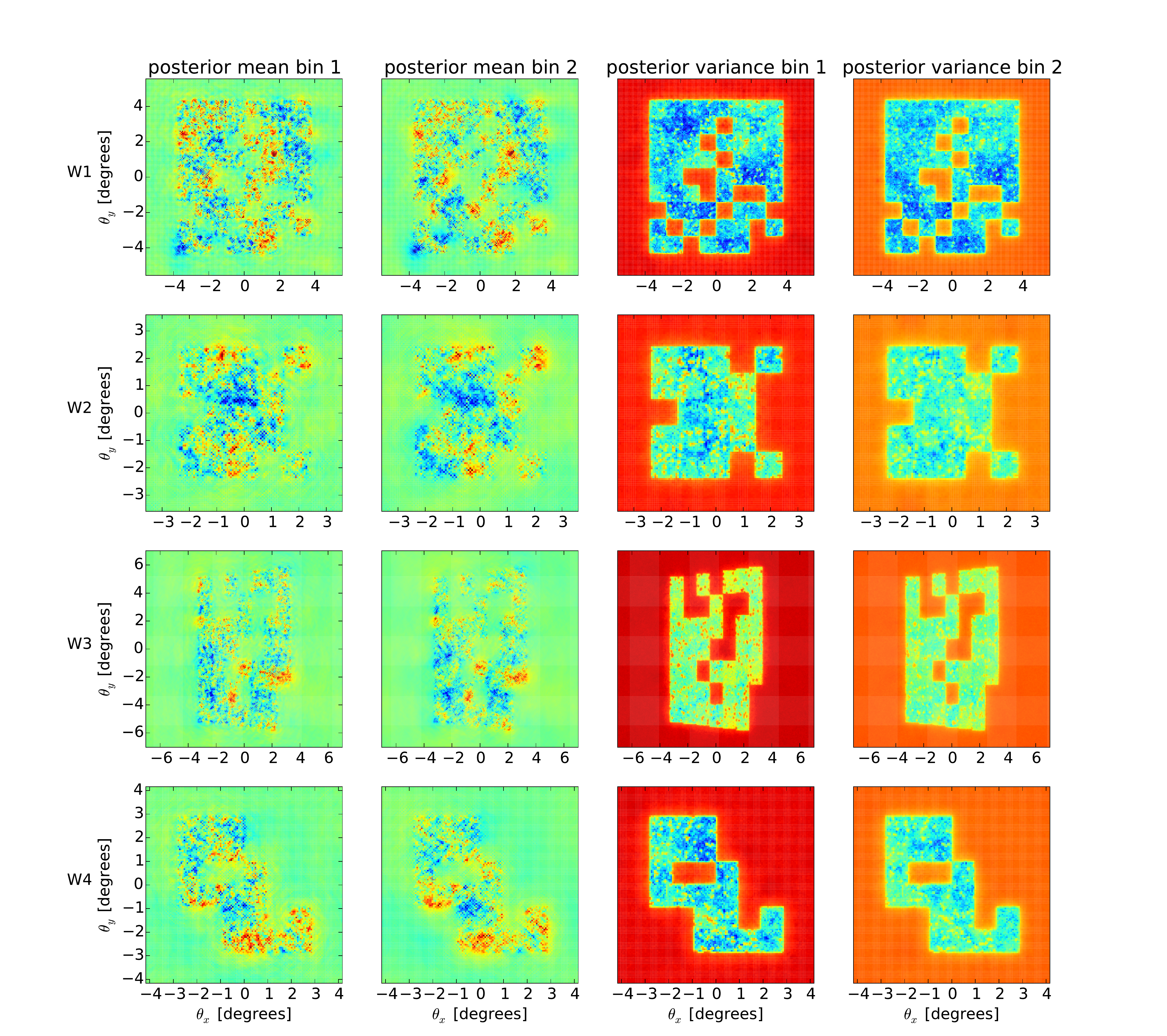}
\includegraphics[width = 17.5cm]{cfhtlens_map_inference_colorbar_small.pdf}
\caption{Recovered posterior mean and variance for the $\gamma_2$ maps in the four CFHTLenS fields and two tomographic bins.}
\label{fig:cfhtlens_map2}
\end{figure*}
The recovered shear maps for the four CFHTLenS fields are shown in Fig. \ref{fig:cfhtlens_map1}-\ref{fig:cfhtlens_map2} -- these figures show the posterior means and variances for the $\gamma_1$ and $\gamma_2$ components respectively. For the first time, we are able to obtain full posterior inference of shear maps from a weak lensing survey. Furthermore, the inferred maps are cosmology independent (notwithstanding the band-power approximations that can be straightforwardly lifted in future analyses) and formally marginalised over our \emph{a priori} uncertainty in the shear power spectrum. The Bayesian inference schemes implemented in this work can also be used to obtain tighter, cosmology-dependent inference of the weak lensing fields by simply assuming a prior on the cosmological parameters or shear power spectra. Such an analysis would formally marginalise over any remaining prior uncertainties on the cosmological parameters. 

Note that the posterior means shown in Fig. \ref{fig:cfhtlens_map1}-\ref{fig:cfhtlens_map2} might not be optimal (or even unbiased) estimators of the true shear maps. For example, the recovered posterior mean map in the masked regions is visibly oversmoothed -- this is expected, since the posterior mean can be thought of as a combination of a large number of Wiener filtered maps, which in the low S/N regime are expected to be oversmoothed and in the limit of infinite noise (\ie, mask) are completely suppressed \citep{Alsing2016}. We emphasize that future scientific applications exploiting shear map inferences should use the full posterior distributions of the shear or convergence in each pixel, or develop a well-motivated estimator from the posterior samples.

The posterior variance in the masked regions (\cf, Fig. \ref{fig:cfhtlens_mask}) is higher than in unmasked regions, as expected. However, the variation of the posterior-variance across the fields is surprisingly small. This can be understood as follows: The shear maps are assumed to be zero mean fields with covariance $\cov$. Since the total variance of the data is given by the signal plus noise covariances $\cov + \noise$ and the noise covariance is known (and fixed), the sample covariance of the data puts a reasonably hard upper limit on the signal covariance. This in turn puts an upper limit on the variance of the true underlying shear field, and hence on the posterior variance of the inferred fields. For the CFHTLenS data the signal-to-noise per pixel is very low across the whole map and the shear variance per pixel is close to saturated everywhere; hence the posterior variance in masked regions where almost all of the information is coming from the prior may only be moderately higher than in unmasked regions where the data are still relatively uninformative. Recall that since the power spectrum is formally marginalised over in the map inference $P(\field | \data)$, the prior on the map $P(\field | \cov)$ at each sampling step ultimately has no impact on the inferred shear maps, other than through imposing the assumption of Gaussian (zero mean) fields. The (hyper) prior on the power spectrum $P(\cov)$ (which is marginalised over), on the other hand, will have some impact on the inferred shear fields and a maximally agnostic analysis should take a reference prior for the power spectrum \citep{Daniels1999}.

\subsection{CFHTLenS cosmological parameters}
\label{sec:cfhtlens_cosmology}
We infer cosmological parameters from the CFHTLenS data using the map-cosmology Gibbs sampling scheme for the three models described in \S \ref{sec:cosmo_models}: a baseline flat-$\Lambda$CDM and two extensions including total neutrino mass and photo-$z$ bias parameters as additional free parameters respectively. For each model, we ran ten chains of $\num[group-separator={,}]{100000}$ steps, ensuring convergence using the usual Gelman-Rubin test $R < 1.03$. Recovered cosmological parameters are summarised in Table \ref{tab:cosmological_parameters} and Fig. \ref{fig:lcdm_cosmology}-\ref{fig:planck_aleviation_z}, and Bayesian model comparison results are summarised in Table \ref{tab:odds_ratios}.

\subsubsection{Baseline flat-$\Lambda$CDM model}
\begin{figure*}
\includegraphics[width = 17.5cm]{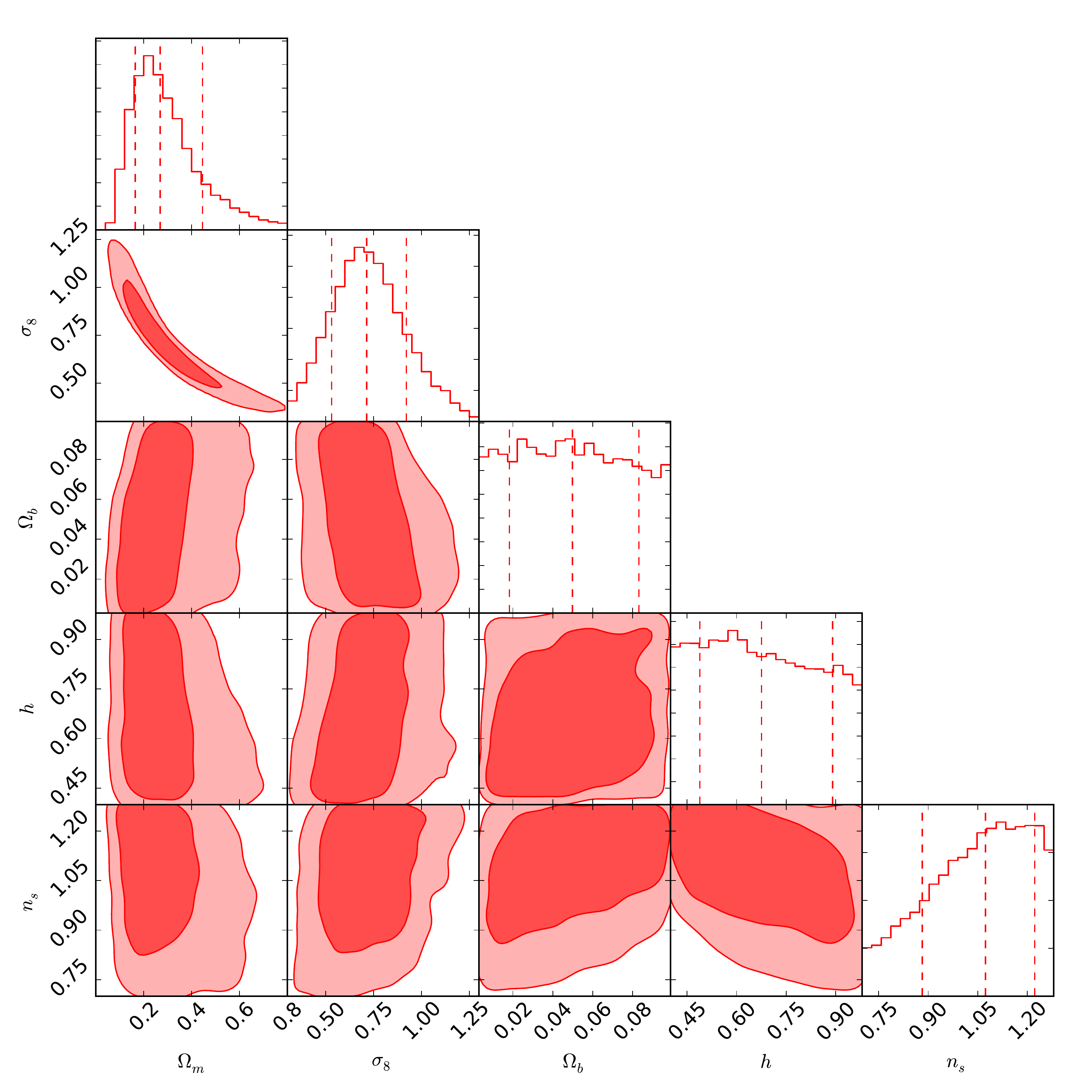}
\caption{Recovered $1$D and $2$D marginal posteriors for the five parameter flat-$\Lambda$CDM model from CFHTLenS, using the joint map-cosmology inference scheme. The contours of the 2D marginals represent $68$ and $95\%$ credible regions respectively, and the dashed lines of the 1D marginals indicate the $16$th, $50$th and $84$th percentiles. }
\label{fig:lcdm_cosmology}
\end{figure*}
The recovered cosmological parameters for the baseline flat-$\Lambda$CDM model are shown in Fig. \ref{fig:lcdm_cosmology}\footnote{Fig. \ref{fig:lcdm_cosmology}, \ref{fig:lcdmnu_cosmology} and \ref{fig:lcdmz_cosmology} were made using a customized version of \textsc{corner.py} \citep{Foreman2014}.} and summarized in Table \ref{tab:cosmological_parameters}. The contours represent $68$ and $95\%$ credible intervals and the vertical dashed lines on the 1D marginals indicate $16th$, $50th$ and $84th$ percentiles. The lensing power spectrum is most sensitive to the scalar amplitude and matter density but with substantial degeneracy between the two, whereas the combination $S_8 =\sigma_8(\Omega_\mathrm{m}/0.3)^{0.5}$ is more strongly constrained (following \citealp{Abbott2015}). Meanwhile, $\Omega_\mathrm{b}$, $h$ and $n_s$ are almost unconstrained by the CFHTLenS data with the scale cuts and two broad tomographic bins taken in this analysis.

\begin{figure*}
\includegraphics[width = 17.5cm]{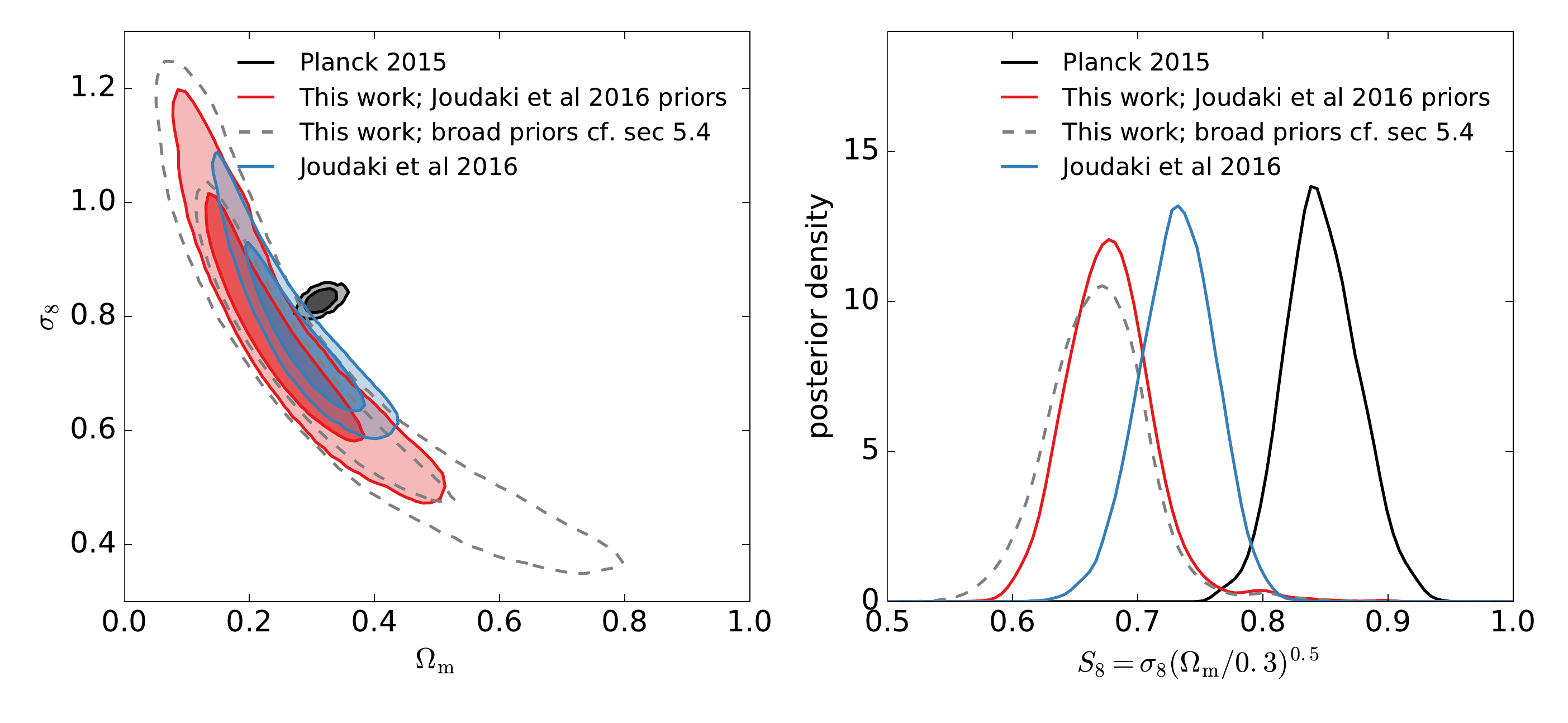}
\caption{Comparison of constraints in $\sigma_8$-$\Omega_\mathrm{m}$ (left) and $S_8 = \sigma_8(\Omega_\mathrm{m}/0.3)^{0.5}$ (right) for the previous CFHTLenS analysis of \citet{Joudaki2016} ($7$-bin tomography; blue), \emph{Planck} 2015 (black) and the present work ($2$-bin tomography) under the broad flat priors described in \S \ref{sec:priors} (grey dashed) and under the same prior assumptions\textsuperscript{\ref{jp}} as \citet{Joudaki2016} (red). The contours in the left panel show 68 and 95\% credible regions.}
\label{fig:cfhtlens_comparison}
\end{figure*}

In Fig. \ref{fig:cfhtlens_comparison} we compare our constraints on $(\sigma_8, \Omega_\mathrm{m})$ and on the combination $S_8 =\sigma_8(\Omega_\mathrm{m}/0.3)^{0.5}$ to the previous 7-bin correlation function analysis of CFHTLenS \citep{Joudaki2016} and to \emph{Planck} 2015. Our results are consistent with \citet{Joudaki2016}, and under the same prior assumptions\footnote{\label{jp}\citet{Joudaki2016} assume a narrower flat priors on the scalar amplitude $\ln A_\mathrm{S}\in [2.3, 5.0]$, Hubble parameter $h\in [0.61, 0.81]$ and Baryon density $\Omega_\mathrm{b}h^2\in[0.013, 0.033]$.} the constraining power of the analyses is comparable; \citet{Joudaki2016} estimate correlation functions in seven tomographic redshift bins and binned at seven angular scales, whereas our analysis splits the data into only two tomographic bins but makes more complete use of angular information, making no binning of angular scales.

Fig. \ref{fig:cfhtlens_comparison} shows that both \citet{Joudaki2016} and our work are in tension with \emph{Planck} 2015 at the level of $\sim$ $2\sigma$, with our results showing slightly worse tension with \emph{Planck} than \citet{Joudaki2016}. Tension between \emph{Planck} and CFHTLenS in the $\sigma_8$-$\Omega_\mathrm{m}$ plane under the flat-$\Lambda$CDM model have been widely reported in previous CFHTLenS analyses, with \citet{Joudaki2016} (7-bin correlation function), \citet{Heymans2013} (6-bin correlation function), \citet{Kitching20143D} (3D power spectrum), \citet{Benjamin2013} (2-bin correlation function), and \citet{Kilbinger2013} (1-bin correlation function) all reporting some level of tension. This tension is interesting because it could indicate evidence for extensions to the baseline-$\Lambda$CDM model, or unaccounted for systematics in the CFHTLenS and/or \emph{Planck} data (both of which are of interest for future cosmological analyses).

Many attempts have been made to explain and/or alleviate this tension. \citet{Joudaki2016} report that the discordance can be alleviated by marginalising over three additional systematic effects in the weak lensing analysis -- intrinsic alignments, baryonic suppression on the small-scale matter power spectrum and photo-$z$ biases -- with reasonably broad priors on all three effects (where none of the systematics were individually able to relieve the tension), but the flexible systematics model is disfavoured by the CFHTLenS data. \citet{MacCrann2015} report that inclusion of an additional (sterile) neutrino is able to alleviate the tension, although this more flexible model is disfavoured by the data, whilst including a massive neutrino or baryonic suppression does little to relieve the discordance. Meanwhile, \citet{Kitching20143D} and \citet{Kohlinger2016} find that aggressively cutting small scales from the weak lensing analysis brings the resulting weak lensing constraints (with substantially larger error bars) into agreement with \emph{Planck}. There may be residual systematic effects in the \emph{Planck} data, too, that are responsible for at least some of the tension (see \eg, \citealp{Addison2015, Spergel2015}). Meanwhile, recent cosmic shear analysis of the Dark Energy Survey (DES) science verification data is consistent with both CFHTLenS and \emph{Planck} \citep{Abbott2015} (with $\sim30\%$ larger error bars compared to CFHTLenS), whilst constraints from the Kilo Degree Survey (KiDS) cosmic shear are consistent with CFHTLenS and in similar $\sim 2\sigma$ tension with \emph{Planck} \citep{Hildebrandt2016}. At this time the jury is still out on the source of the tension between \emph{Planck} and CFHTLenS weak lensing.
\begin{table*}
\small
\begin{center}
\caption{Marginal parameter constraints on the cosmological parameters from CFHTLenS for the three models considered: baseline (flat) $\Lambda$CDM and two extensions including total neutrino mass and photo-$z$ bias parameters as additional free parameters respectively. The maximum posterior values and 68\% credible intervals are given. The quantity $S_8$ is defined as $S_8 = \sigma_8(\Omega_\mathrm{m}/0.3)^{0.5}$ and the ``informative" prior for the $\Lambda$CDM$+\Delta z$ model refers to the Gaussian prior on the photo-$z$ bias parameters derived from the CFHTLenS-BOSS cross-correlation analysis of \citet{Choi2015} (\cf, Table \ref{tab:choi}). Flat priors are defined in \S \ref{sec:priors}.}
\begin{tabular}{ccccccccc}
\hline
Model & Prior & $\sigma_8$ & $S_8$ & $\ln(10^{10}A_\mathrm{S})$ & $\Omega_\mathrm{m}$ &  $\sum m_\nu$ & $p_1$ & $p_2$\\
\hline\tblspacer\\
$\Lambda$CDM & Flat & $ 0.69 ^{+ 0.23 }_{- 0.17 }$ & $ 0.67 ^{+ 0.03 }_{- 0.03 }$ & $ 2.89 ^{+ 1.20 }_{- 1.14 }$ & $ 0.23 ^{+ 0.14 }_{- 0.10 }$  & - & - & -\tblspacer \\ 
$\Lambda$CDM+$m_\nu$ & Flat & $ 0.60 ^{+ 0.20 }_{- 0.15 }$ & $ 0.67 ^{+ 0.04 }_{- 0.04 }$ & $ 2.79 ^{+ 1.41 }_{- 1.23 }$ & $ 0.26 ^{+ 0.19 }_{- 0.12 }$ & $< 4.6 (95\%)$& - & -\tblspacer \\ 
$\Lambda$CDM+$\Delta z$ &Flat & $ 0.63 ^{+ 0.28 }_{- 0.09 }$ &$ 0.70 ^{+ 0.16 }_{- 0.12 }$ &$ 2.38 ^{+ 1.50 }_{- 0.75 }$ &$ 0.20 ^{+ 0.22 }_{- 0.09 }$ &- &$ -0.25 ^{+ 0.53 }_{- 0.60 }$ &$ -0.15 ^{+ 0.17 }_{- 0.15 }$ \tblspacer \\ 
$\Lambda$CDM+$\Delta z$ & Informative & $ 0.70 ^{+ 0.18 }_{- 0.13 }$ &$ 0.70 ^{+ 0.03 }_{- 0.03 }$ &$ 2.80 ^{+ 1.22 }_{- 1.22 }$ &$ 0.24 ^{+ 0.16 }_{- 0.09 }$ & - &$ 0.45 ^{+ 0.02 }_{- 0.03 }$ &$ -0.17 ^{+ 0.03 }_{- 0.01 }$ \tblspacer \\ 
\hline
\end{tabular}
\label{tab:cosmological_parameters}
\end{center}
\end{table*}
\begin{table}
\small
\begin{center}
\caption{Model comparison of the baseline $\Lambda$CDM and two extended models, with three degenerate massive neutrinos and (redshfit dependent) photo-$z$ biases respectively. The log Bayes factor $2\ln\mathcal{K}$ is given for $\Lambda$CDM versus the extended model, and we adopt the quantitative Bayes factor interpretation scheme of \citet{Kass1995}. }
\begin{tabular}{cccc}
\hline
Model & Prior & $2\;\ln\mathcal{K}$ & Interpretation \\
\hline\tblspacer \\
$\Lambda$CDM+$m_\nu$ & Flat & -0.3 & Support for $\Lambda$CDM  \\ 
$\Lambda$CDM+$\Delta z$ & Flat & -0.45 & Support for $\Lambda$CDM  \\
$\Lambda$CDM+$\Delta z$ & Informative & -0.73 & Support for $\Lambda$CDM \tblspacer \\ 
\hline
\end{tabular}
\label{tab:odds_ratios}
\end{center}
\end{table}
\subsubsection{Massive neutrinos}
\begin{figure*}
\includegraphics[width = 17.5cm]{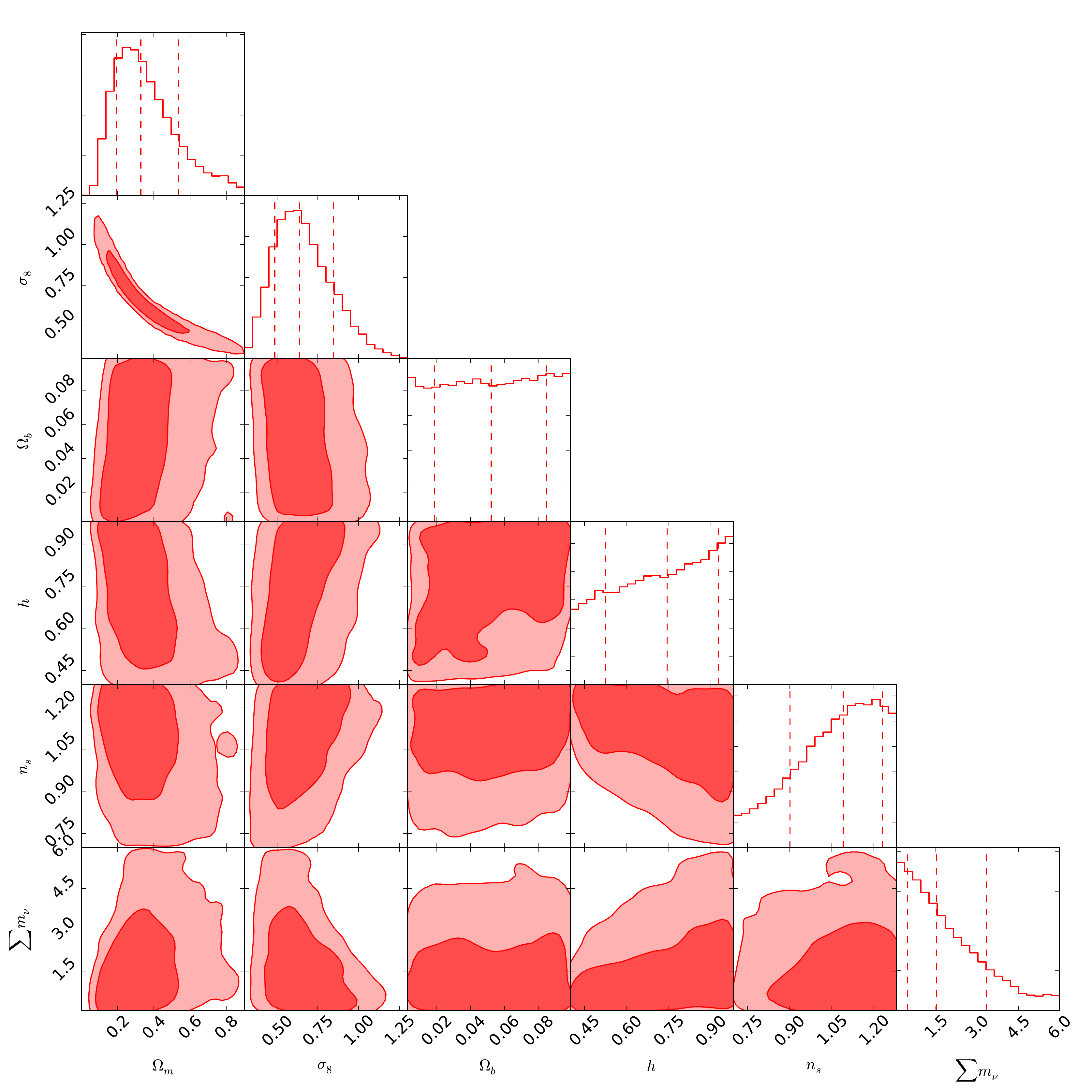}
\caption{Recovered $1$D and $2$D marginal posteriors for the $\Lambda$CDM$+m_\nu$ model from CFHTLenS, with three degenerate massive neutrinos of total mass $\sum m_\nu$. The contours of the 2D marginals represent $68$ and $95\%$ credible regions respectively, and the dashed lines of the 1D marginals indicate the $16$th, $50$th and $84$th percentiles. The CFHTLenS data constrains the total neutrino mass to $\sum m_\nu < 4.6\mathrm{eV}$ at 95\% credibility.}
\label{fig:lcdmnu_cosmology}
\end{figure*}

Fig. \ref{fig:lcdmnu_cosmology} shows the recovered cosmological parameters for the $\Lambda$CDM$+m_\nu$ model including three degenerate massive neutrinos with total mass $\sum m_\nu$. From the CFHTLenS data alone we obtain a weak upper limit on the total neutrino mass of $\sum m_\nu < 4.6\mathrm{eV}$ ($95\%$). In comparison, \emph{Planck} 2015 report an upper limit of $\sum m_\nu < 0.23\mathrm{eV}$ (95\%) from CMB observations alone, and combined CMB and BAO measurements constrain $\sum m_\nu < 0.17\mathrm{eV}$.

\begin{figure}
\includegraphics[width = 8cm]{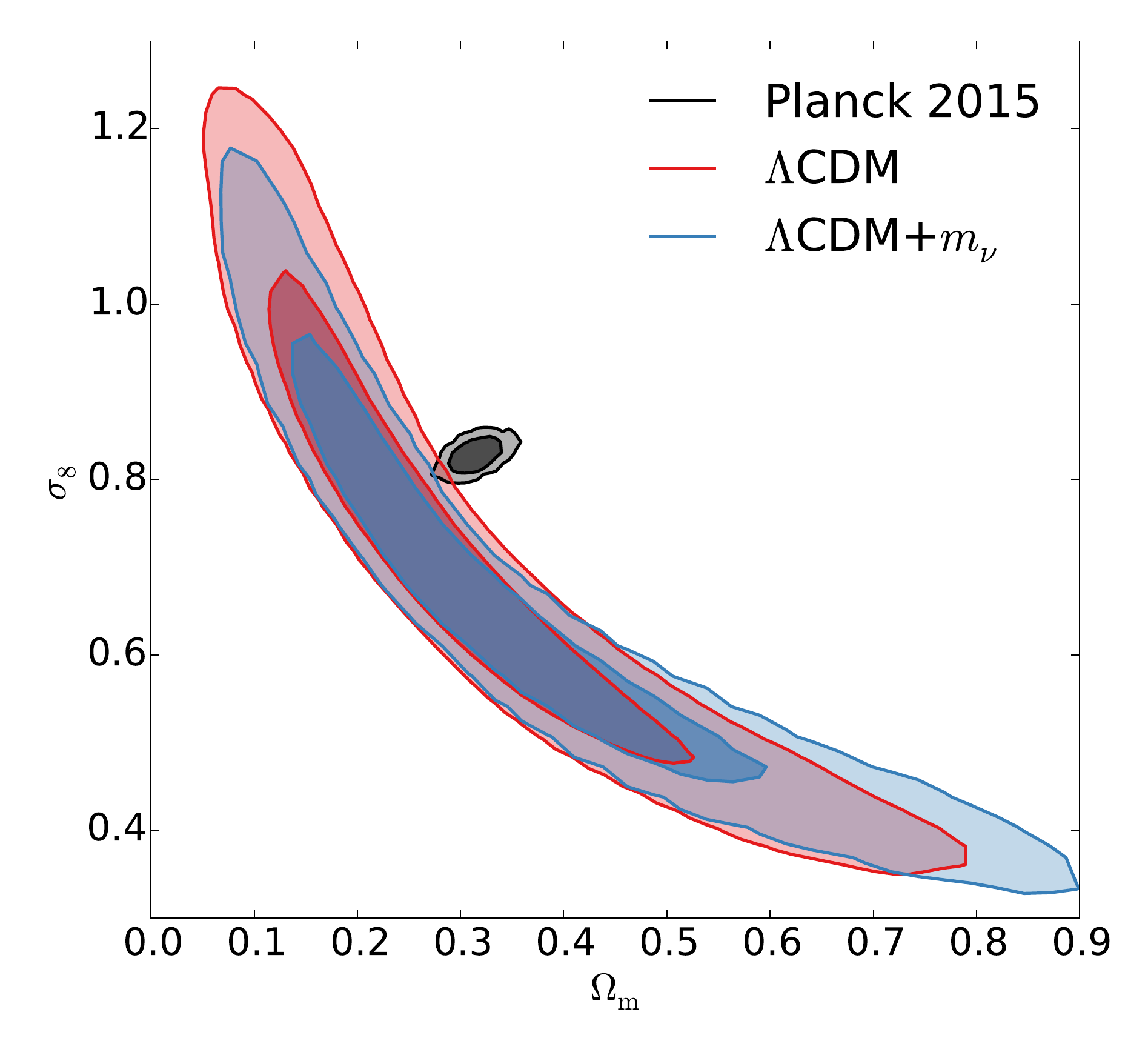}
\caption{Effect of adding total neutrino mass $\sum m_\nu$ as an additional free parameter on the $(\sigma_8, \Omega_\mathrm{m})$ constraints from CFHTLenS and tension with Planck; the tension is not alleviated by the addition of neutrino mass as a free parameter. }
\label{fig:planck_aleviation_nu}
\end{figure}

In Fig. \ref{fig:planck_aleviation_nu} we see that including massive neutrino masses does little to alleviate tensions with Planck, having a relatively modest impact on the $(\sigma_8, \Omega_\mathrm{m})$ inferences. The log Bayes factor for $\Lambda$CDM$+m_\nu$ versus $\Lambda$CDM is $2\;\ln\mathcal{K} = -0.3$, weakly supporting $\Lambda$CDM; the CFHTLenS data alone do not prefer a non-minimal neutrino mass (Table \ref{tab:odds_ratios}).

When computing the Bayes factor for $\Lambda$CDM$+m_\nu$ versus $\Lambda$CDM in Table \ref{tab:odds_ratios}, we approximate the two models as being nested. Strictly speaking the model with three massive neutrinos does not quite map continuously on to the baseline model (with only one massive neutrino). However, the difference in the tomographic power spectra between the baseline model and the degenerate neutrino model with $\sum m_\nu = 0.06$ is small; within the constraining power of CFHTLenS, this difference is negligible and under a flat prior on the $\Lambda$CDM parameters the Savage Dickey Density Ratio is a good approximation for the Bayes factor.
\subsubsection{Photo-$z$ bias}
\begin{figure*}
\includegraphics[width = 17.5cm]{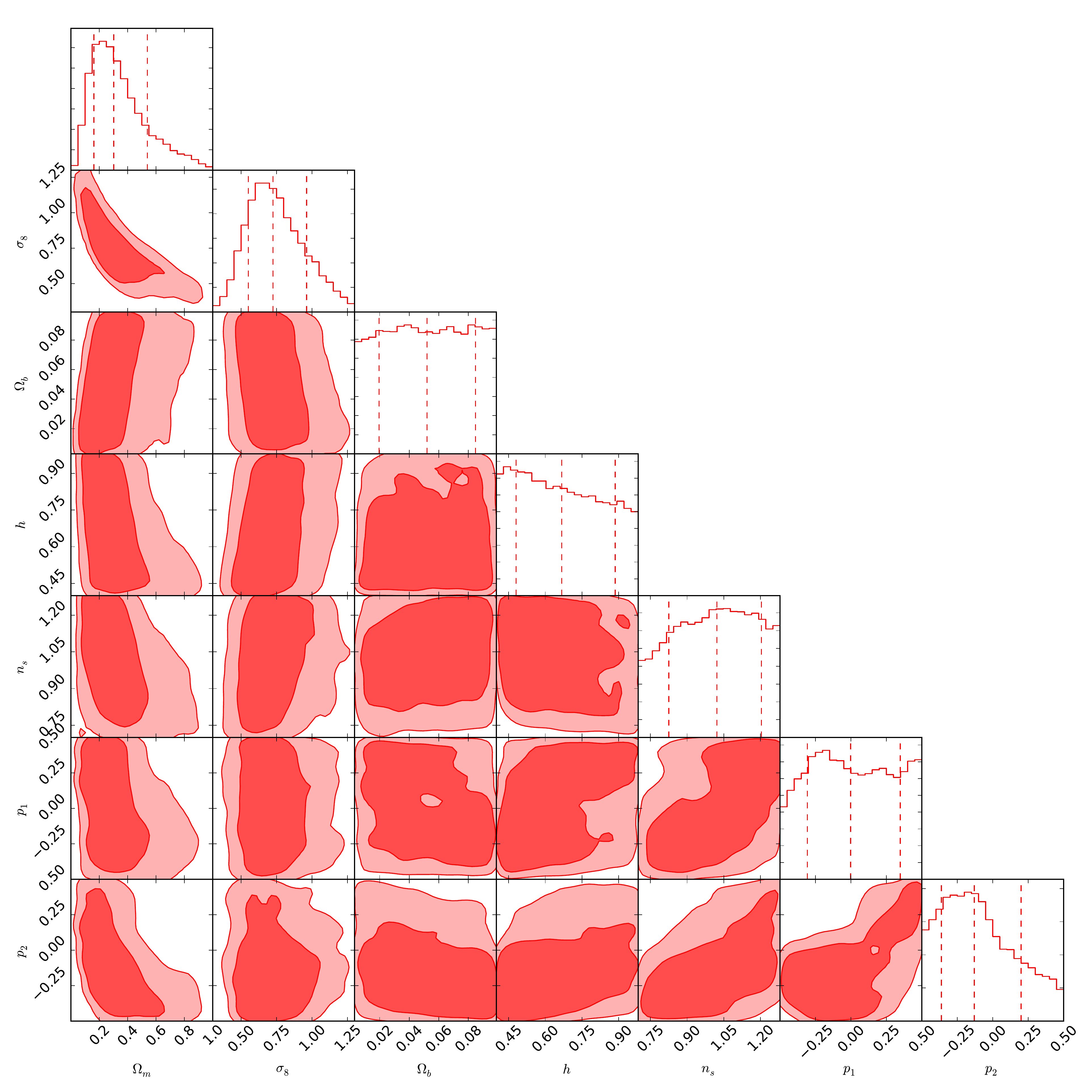}
\caption{Recovered $1$D and $2$D marginal posteriors for the $\Lambda$CDM$+\Delta z$ model from CFHTLenS, assuming a linear redshift dependent photo-$z$ bias parametrized by $\Delta z = p_2(z - p_1)$. The contours of the 2D marginals represent $68$ and $95\%$ credible regions respectively, and the dashed lines of the 1D marginals indicate the $16$th, $50$th and $84$th percentiles. }
\label{fig:lcdmz_cosmology}
\end{figure*}

Fig. \ref{fig:lcdmz_cosmology} shows the recovered cosmological parameters for the $\Lambda$CDM$+\Delta z$ model including a linear bias model $\Delta z = p_2(z - p_1)$ for the photo-$z$ (described in \S \ref{sec:cosmo_models}), following indications from previous studies that CFHTLenS may suffer from redshift dependent photo-$z$ bias \citep{Choi2015, Kitching2016, Joudaki2016}. Using the CFHTLenS data alone, we find that $p_1=-0.25 ^{+ 0.53 }_{- 0.60 }$ and $p_2 = -0.15 ^{+ 0.17 }_{- 0.15 }$. Both parameters are consistent with zero. For comparison, \citet{Kitching2016} (assuming a fixed \emph{Planck} cosmology) found $p_1=0.26 \pm 0.05$ and $p_2 = -0.25 \pm 0.06$, and \citet{Choi2015} found $p_1=0.45 \pm 0.05$ and $p_2 = -0.16 \pm 0.05$. Whilst the broad constraints on $p_1$ and $p_2$ from the CFHTLenS data alone are consistent with both previous studies, it appears that the CFHTLenS data, assumption of a \emph{Planck} cosmology and cross-correlation between CFHTLenS and BOSS are pulling the photo-$z$ parameters in somewhat different directions; notably, the constraints from \citet{Choi2015} and the values found by \citet{Kitching2016} required to bring about concordance with \emph{Planck} are in tension.

\begin{figure*}
\includegraphics[width = 18cm]{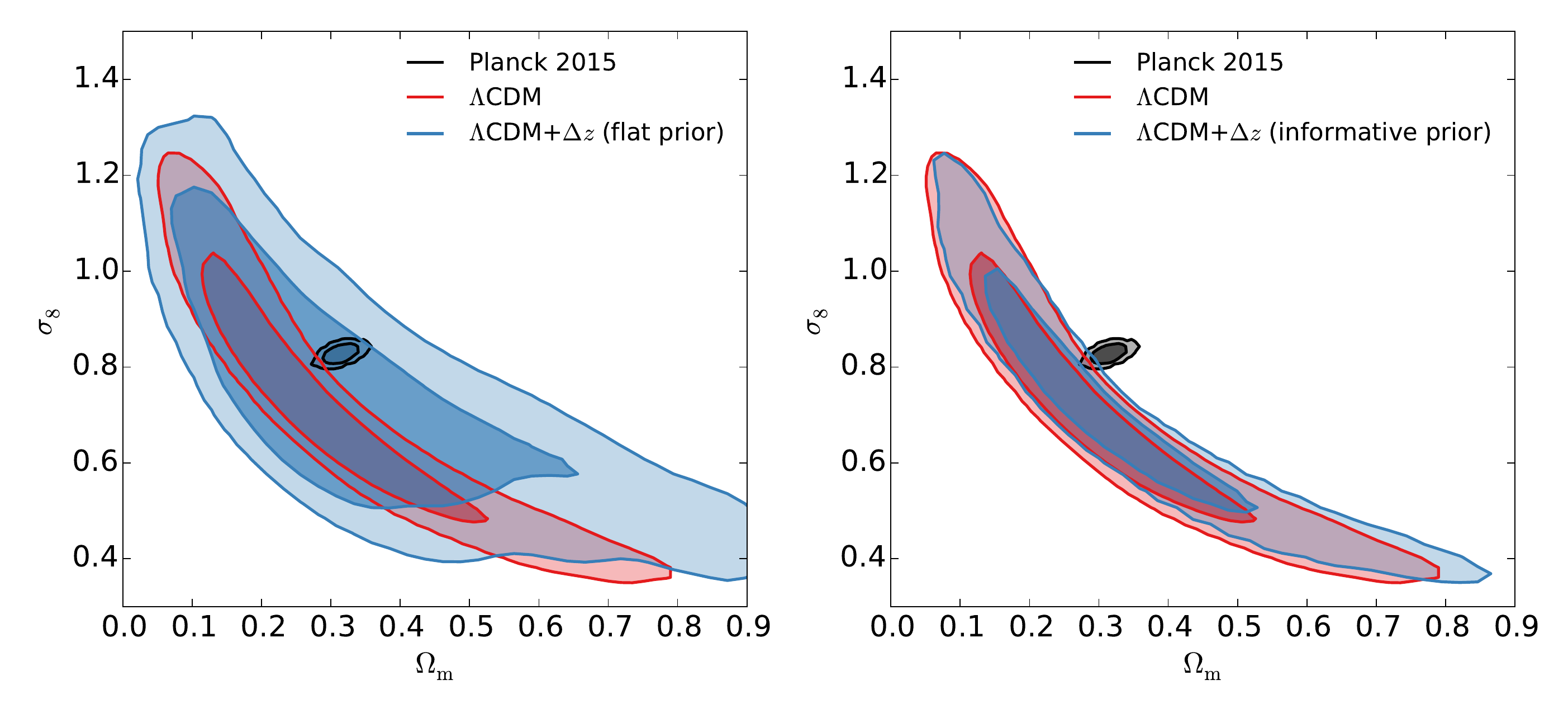}
\caption{Effect of adding linear (redshift dependent) photo-$z$ bias on the $(\sigma_8, \Omega_\mathrm{m})$ constraints from CFHTLenS and tension with Planck. Under a broad uniform prior on the photo-$z$ bias parameters (left), the constraints are significantly degraded and tension between CFHTLenS and Planck is alleviated through an increase in the error bars from CFHTLenS. Meanwhile, under an informative prior on the photo-$z$ bias parameters from \citet{Choi2015} (right panel) the photo-$z$ bias has a modest impact on the tension between CFHTLenS and Planck. }
\label{fig:planck_aleviation_z}
\end{figure*}

In Fig. \ref{fig:planck_aleviation_z} we see that under a broad flat prior (left panel), the addition of the two additional photo-$z$ bias parameters alleviates tension with \emph{Planck} through a substantial increase in the error bars. Meanwhile, under the informative Gaussian prior on $(p_1, p_2)$ as derived from the CFHTLenS-BOSS cross correlation analysis of \citep{Choi2015} (\cf, \S \ref{sec:priors}), the modification to the $(\Omega_\mathrm{m}, \sigma_8)$-constraints is more modest; the contours are shifted slightly towards \emph{Planck} but the tension remains. Comparing the log Bayes factor for $\Lambda$CDM$+\Delta z$ versus $\Lambda$CDM we find that the CFHTLenS data alone do not prefer the $\Lambda$CDM$+\Delta z$ model under either the flat or informative prior (Table \ref{tab:odds_ratios}). In fact, the photo-$z$ bias model is slightly more disfavoured under the informative prior than the flat prior; this is due to the fact that the photo-$z$ bias values found by \citet{Choi2015} fall in a region of relatively low posterior density from our analysis of the CFHTLenS data alone (\cf, Fig. \ref{fig:lcdmz_cosmology}). Lack of preference for $\Lambda$CDM$+\Delta z$ over the baseline model is consistent with the analysis in \citet{Joudaki2016}.

\section{Computational cost and prospects for future large-area surveys}
\label{sec:cost}
In this section we consider the computational cost of the Bayesian inference schemes relative to the traditional approach of sampling cosmological parameters from a likelihood built from an estimator for the 2-point statistics (\ie, the power spectrum or correlation function). We consider a future large-area survey analysis performed on the full curved-sky (\ie, with Fourier transforms replaced with spherical harmonic transforms) with $10$ tomographic bins and a resolution of $\ell_\mathrm{max} = 4096$. We will argue that the Bayesian schemes are expected to be of comparable cost to traditional methods and thus represent a practical approach for upcoming large-area surveys such as \emph{Euclid} and \emph{LSST}. 

\subsubsection*{Map-cosmology sampling}

When comparing the relative cost of two MCMC sampling methods there are two key considerations: cost per MCMC sample and the MCMC correlation length. The total cost per \emph{independent} sample is the cost per sample multiplied by the correlation length.

\citet{Racine2015} demonstrated that the map-cosmology inference scheme provides MCMC chains with comparable correlation lengths (within a factor of $2$) compared to running state-of-the-art cosmology samplers (\eg, \textsc{cosmoMC}; \citealp{Lewis2002}) on the \emph{Planck} likelihood. Since the sampling scheme presented here and in \citet{Racine2015} is designed to be (roughly) invariant with respect to signal-to-noise, we also expect comparable correlation lengths for weak lensing applications (although we do not attempt a direct comparison here). With comparable correlation lengths, the remaining main difference in cost between the map-cosmology sampler and traditional estimator-likelihood sampling is the cost per MCMC sample. 

When sampling estimator-likelihoods, the cost per sample is dominated by computing the lensing power spectra at each step (via the non-linear matter power spectrum). For the current state-of-the-art weak lensing analysis code \textsc{cosmosis} \citep{Zuntz2015} operating at tolerances required for the Dark Energy Survey (DES), computing tomographic power spectra for an $n_\mathrm{bins} = 10$ bin analysis costs $\sim 50$ CPU seconds on a high-end 2016 CPU. For the map-cosmology inference scheme, there are two main costs per MCMC cycle: computing the lensing power spectra, which will be the same as for any other approach, and performing the $2\times n_\mathrm{bins}$ spherical harmonic transforms (SHT) at the map and messenger field sampling steps. The current state-of-the-art SHT \citep{Reinecke2013} cost $\sim 1$ CPU second per SHT at a resolution of $\ell_\mathrm{max} = 4096$ on a high-end 2016 CPU, so the total cost of spherical harmonic transforms per MCMC cycle is hence $\sim 20$ CPU seconds. Therefore, the total cost per MCMC sample for the map-cosmology sampler and traditional likelihood sampling will be similar to within a factor of $2$ or so. Computing lensing power spectra and spherical harmonics both parallelise very well.

With the expectation of comparable correlation lengths (to within a factor of $2$), and comparable cost per MCMC sample (also within a factor of $2$), we expect that the total cost of the Bayesian map-cosmology sampling scheme will be comparable to traditional estimator-likelihood sampling methods such as \textsc{cosmoMC} to within a modest factor of $2$--$4$.

\subsubsection*{Map-power spectrum sampling}
In a full curved-sky analysis, the cost per MCMC sample for map-power spectrum sampling will be dominated by the $2\times n_\mathrm{bins}$ spherical harmonic transforms required per Gibbs cycle. As discussed above, with current technology this costs in the region of $20$ CPU seconds per Gibbs sample for an $n_\mathrm{bins} = 10$ tomographic bin analysis. With a modest amount of parallelisation, \eg, using $2\times n_\mathrm{bins}$ cores, $\sim 10^6$ samples should be achievable on the timescale of $\sim 10$ days (with current technology). The total number of samples required for reconstructing the power spectrum posterior for future large-area surveys is not precisely known. However, we expect the signal-to-noise per $\ell$ mode for \emph{Euclid} to be comparable to the signal-to-noise per band-power considered in this work for CFHTLenS (with $\Delta \ell = 200$ but only $1/100$th of the survey area). We found MCMC convergence could be achieved with $\sim 10^6$ samples, so this may be a representative number for future surveys.

Note that the joint sampling step of \citet{Racine2015} applied here to map-cosmology sampling can equally well be applied to map-power spectrum inference; we expect this sophistication to result in a dramatic improvement in the total cost of map-power spectrum sampling by reducing the correlation length of the MCMC chains in the low signal-to-noise regime (typical for lensing).

In summary, both the map-cosmology and map-power spectrum sampling schemes are comfortably computationally feasible for current and future weak lensing survey analyses.

\section{Conclusions}
\label{sec:conclusions}
We have applied Bayesian map-power spectrum and map-cosmology hierarchical inference schemes for extracting cosmic shear power spectra, shear maps and cosmological parameters from CFHTLenS -- the first application to data. Under the baseline flat-$\Lambda$CDM model, we obtain cosmological parameter constraints consistent with previous CFHTLenS analyses, and in-line with previous studies our inferred $(\sigma_8, \Omega_\mathrm{m})$ constraints are in tension with \emph{Planck} 2015 results at the $2\sigma$ level (under the baseline $\Lambda$CDM model). 

Extending the baseline model to include massive neutrinos, we are able to constrain the total neutrino mass to $\sum m_\nu < 4.6\mathrm{eV}$ (95\%) from CFHTLenS data alone. The inclusion of neutrino mass as an extra degree-of-freedom does little to alleviate the tension between CFHTLenS and Planck, and the more flexible model is not preferred over $\Lambda$CDM by the CFHTLenS data.

Including the possibility of a linear redshift-dependent photo-$z$ bias $\Delta z = p_2(z - p_1)$ we find the CFHTLenS data prefer $p_1=-0.25 ^{\scriptscriptstyle+ 0.53 }_{\scriptscriptstyle- 0.60 }$ and $p_2 =  -0.15 ^{\scriptscriptstyle+ 0.17 }_{\scriptscriptstyle- 0.15 }$, although both are consistent with zero. Including $p_1$ and $p_2$ as additional parameters under a broad flat prior completely alleviates tension between CFHTLenS and Planck in our analysis, due to a significant increase in the error bars from the extra degrees-of-freedom. Imposing an informative prior on $(p_1, p_2)$ from the CFHTLenS-BOSS cross-correlation analysis of \citet{Choi2015}, the impact of the photo-$z$ bias on the cosmological constraints is more modest and the tension with Planck remains. The CFHTLenS data alone do not prefer the more flexible photo-$z$ bias model over the baseline $\Lambda$CDM under either the flat or informative prior.

As well as demonstrating the Bayesian inference schemes on current data, we have argued that for future large-area surveys such as \emph{Euclid} and \emph{LSST} the map-cosmology sampling scheme is of comparable computational cost to traditional estimator-likelihood sampling methods, and the map-power spectrum sampling scheme is also a computationally practical approach for obtaining cosmology independent power spectrum inference for future surveys. 

Both the map-power spectrum and map-cosmology inference schemes implemented in this work assume Gaussian lensing fields. Whilst this is appropriate (and optimal) on large scales, on smaller scales the lensing fields are well known to be non-Gaussian. This is both an opportunity and a curse: if we are to assume Gaussianity, we must rigorously validate the algorithms against $N$-body/non-Gaussian shear simulations to test for any model biases that could arise from the Gaussian assumption. On the other hand, the Bayesian hierarchical approach can be readily extended to include models for the non-Gaussian shear, allowing us to extract information beyond the two-point statistics and exploiting the full information content of the cosmological fields, leading to tighter constraints on cosmology and better science at the end of the day. We will explore the limits of the Gaussian approximation for weak lensing analyses and develop extended hierarchical models for non-Gaussian shear inference in future work.


\section*{acknowledgements}
We would like to thank Benjamin Wandelt, Hans-Kristian Eriksen, Benjamin Racine, Fabian K\"{o}linger, Tom Kitching and Till Hoffmann for invaluable discussions. We also thank Benjamin Joachimi for providing the lognormal shear simulations used in this work and Shahab Joudaki for providing MCMC chains for comparison with the previous CFHTLenS analysis.
\bibliographystyle{mn2e_williams}
\bibliography{references}

\end{document}